\documentclass[review]{elsarticle}
\usepackage[english]{babel} 
\usepackage[utf8]{inputenc}
\usepackage{multirow}
\usepackage{acronym}
\usepackage{amsmath,amsfonts, amssymb} 
\usepackage{pifont}
\usepackage{rotating}
\usepackage{url}
\usepackage{soul}
\usepackage{float}
\usepackage{array}
\usepackage{amsmath}
\usepackage{makecell}
\usepackage{balance}
\usepackage{xcolor}
\usepackage{xr}
\usepackage{lscape}
\usepackage{verbatim}
\usepackage{acronym}
\usepackage{graphicx}
\usepackage{adjustbox}
\usepackage{booktabs}
\usepackage{comment}
\usepackage{lineno,hyperref}
\modulolinenumbers[5]

\journal{Computer Networks}

\acrodef{AES}{Advanced Encryption Standard}
\acrodef{AF}{Amplify and Forward}
\acrodef{AGC}{Automatic Gain Control}
\acrodef{AI}{Artificial Intelligence}
\acrodef{AWGN}{Additive White Gaussian Noise}
\acrodef{CDMA}{Code Division Multiple Access}
\acrodef{CNR}{Carrier-to-Noise Ratio}
\acrodef{CPS}{Cyber-Physical Systems}
\acrodef{CRN}{Cognitive Radio Network}
\acrodef{CSI}{Channel State Information}
\acrodef{D2D}{Device-to-Device Communications}
\acrodef{DF}{Decode and Forward}
\acrodef{DoS}{Denial of Service}
\acrodef{ECC}{Elliptic Curve Cryptography}
\acrodef{ECDSA}{Elliptic Curve Digital Signature Algorithm}
\acrodef{ESA}{European Space Agency}
\acrodef{ESR}{Ergodic Secrecy Rate}
\acrodef{FDMA}{Frequency Division Multiple Access}
\acrodef{GG}{Ground-to-Ground}
\acrodef{GNSS}{Global Navigation Satellite System}
\acrodef{GPS}{Global Positioning System}
\acrodef{IRS}{Intelligent Reflecting Surfaces}
\acrodef{IoST}{Internet of Space Things}
\acrodef{IoT}{Internet of Things}
\acrodef{MIMO}{Multiple-Input Multiple-Output}
\acrodef{MISO}{Multiple-Input Single-Output}
\acrodef{MITM}{Man-In-The-Middle}
\acrodef{mMIMO}{Massive \ac{MIMO}}
\acrodef{NFC}{Near Field Communication}
\acrodef{NIC}{Network Interface Card}
\acrodef{NIST}{National Institute of Standards and Technology}
\acrodef{NOAA}{National Oceanic and Atmospheric Administration}
\acrodef{NOMA}{Non-Orthogonal Multiple Access}
\acrodef{OFDM}{Orthogonal Frequency Division Multiplexing}
\acrodef{OFDMA}{Orthogonal Frequency Division Multiplexing Access}
\acrodef{OTP}{One-Time-Pad}
\acrodef{RSA}{Rivest-Shamir-Adleman}
\acrodef{PDR}{Packet Delivery Ratio}
\acrodef{PS}{Power Splitting}
\acrodef{QoS}{Quality of Service}
\acrodef{RF}{Radio Frequency}
\acrodef{RFID}{Radio Frequency IDentification}
\acrodef{RSMA}{Rate-Splitting Multiple Access}
\acrodef{SATCOM}{Satellite-based Communication}
\acrodef{SDN}{Software Defined Networking}
\acrodef{SDR}{Software Defined Radio}
\acrodef{SEE}{Secrecy Energy Efficiency}
\acrodef{SG}{Satellite-to-Ground}
\acrodef{SIMO}{Single-Input Multiple-Output}
\acrodef{SINR}{Signal to Noise plus Interference Ratio}
\acrodef{SISO}{Single-Input Single-Output}
\acrodef{SNR}{Signal-to-Noise Ratio}
\acrodef{SOP}{Secrecy Outage Probability}
\acrodef{SRM}{Secrecy Rate Maximization}
\acrodef{SS}{Satellite-to-Satellite}
\acrodef{TDMA}{Time Division Multiple Access}
\acrodef{TS}{Time Splitting}
\acrodef{UAV}{Unmanned Aerial Vehicles}
\acrodef{VLC}{Visible Light Communications}
\acrodef{VSAT}{Very Small Aperture Terminal}

\newcolumntype{P}[1]{>{\centering\arraybackslash}m{#1}}
\newcommand{\cmark}{\ding{51}}%
\newcommand{\xmark}{\ding{55}}%

\acrodef{ML}{Machine Learning}
\acrodef{AES}{Advanced Encryption Standard}
\acrodef{QKD}{Quantum Key Distribution}
\acrodef{P2P}{Point to Point}

\graphicspath{{figures/}}

\begin{document}
\begin{frontmatter}

\title{Satellite-Based Communications Security: A Survey of Threats, Solutions, and Research Challenges}

\author[1]{Pietro Tedeschi}
\author[2]{Savio Sciancalepore}
\author[3]{Roberto Di Pietro}

\address[1]{Technology Innovation Institute, Autonomous Robotics Research Center, Abu Dhabi, United Arab Emirates\\ e-mail: pietro.tedeschi@tii.ae}

\address[2]{Eindhoven University of Technology, Eindhoven, Netherlands \\ e-mail: s.sciancalepore@tue.nl}

\address[3]{Division of Information and Computing Technology (ICT) \\ College of Science and Engineering (CSE), Hamad Bin Khalifa University (HBKU), Doha, Qatar \\ e-mail: rdipietro@hbku.edu.qa}

\begin{abstract}
Satellite-based Communication (SATCOM) systems are gaining renewed momentum in Industry and Academia, thanks to innovative services introduced by leading tech companies and the promising impact they can deliver towards the \emph{global connectivity} objective tackled by early 6G initiatives. On the one hand, the emergence of new manufacturing processes and radio technologies promises to reduce service costs while guaranteeing outstanding communication latency, available bandwidth, flexibility, and coverage range. On the other hand, cybersecurity techniques and solutions applied in SATCOM links should be updated to reflect the substantial advancements in attacker capabilities characterizing the last two decades. However, business urgency and opportunities are leading operators towards challenging system trade-offs, resulting in  an increased attack surface and a general relaxation of the available security services.

In this paper, we tackle the cited problems and present a comprehensive survey on the link-layer security threats, solutions, and challenges faced when deploying and operating SATCOM systems. Specifically, we classify the literature on security for SATCOM systems into two main branches, i.e., physical-layer security and cryptography schemes. Then, we further identify specific research domains for each of the identified branches, focusing on dedicated security issues, including, e.g., physical-layer confidentiality, anti-jamming schemes, anti-spoofing strategies, and quantum-based key distribution schemes. For each of the above domains, we highlight the most essential techniques, peculiarities, advantages, disadvantages, lessons learned, and future directions.
Finally, we also identify emerging research topics whose additional investigation by Academia and Industry could further attract researchers and investors, ultimately unleashing the full potential behind ubiquitous satellite communications.
\end{abstract}



\end{frontmatter}

\section{Introduction}
\label{sec:intro}

\acp{SATCOM} play a vital role in the global telecommunication systems, having found  applications in a plethora of  domains throughout the last 50 years, including radio broadcasting, weather forecast, maritime communications, assisted navigation, and military operations, to name a few~\cite{maral2020satellite}. 
While the attention of Academia and Industry in the last years was mainly focused on ground communication systems, recent business initiatives launched by leading tech companies such as SpaceX, Facebook, and Amazon generated new renewed interest in satellite-based systems~\cite{spacex_starlink, theverge_facebook_amazon}. In particular, satellites are being deployed to provide services in a variety of new application domains, e.g., to reach remote locations providing unmatched connectivity (as per bandwidth and cost), or to support low-power  constrained \ac{IoT} devices~\cite{fang2021_iotj}. 
As a result, recent commercial and standardization activities clearly indicate \acp{SATCOM} as one of the most important enabling technologies for supporting the development of the upcoming sixth-generation (6G) networks~\cite{Rappaport2019}. 
In addition, the business driving factors also seem to indicate a bright future for \acp{SATCOM}. Indeed, according to a dedicated research report by Market Research Future (MRFR), ``Satellite Communication Market Information by Product, Technology, End-Use, and Region - Forecast till 2025”, the \ac{SATCOM} market is anticipated to reach USD $41,860$ Million by 2025, sporting a  $8.40$\% Compound Annual Growth Rate (CAGR). 

Despite the promising applications and forecasting, the adoption of satellite links generally widens the threat surface of a system, introducing new vulnerabilities. Indeed, the ease to eavesdrop, tamper, disrupt, and reroute the satellite traffic provides the attacker with an extensive portfolio of opportunities to launch cyber-attacks at scale and affect the operations of such systems in different ways~\cite{santamarta_2014_blackhat}. 
To complete the scenario, there is also the fact that the military satellite backbone is subject to the same (if not bigger) issues \cite{sat_under_attack}.

The severity of the cited threats could be even higher when considering that most of the current satellite systems either do not integrate security at all 
or run outdated security techniques, unable to face the complex attacks launched today~\cite{manulis2020cyber}. As a result, security solutions for SATCOM should be revisited 
and coupled with the unique features of satellite-based systems.

Several contributions in the last decade already investigated security issues in the context of \ac{SATCOM} (refer to Table~\ref{tab:related_with_features} for an overview).
\begin{table}[htbp]
\scriptsize
\centering
\caption{Main addressed topics and differences between surveys touching \ac{SATCOM} systems.}
\begin{tabular}{|P{1cm}|P{1.8cm}|P{1.2cm}|P{1.2cm}|P{1.6cm}|P{1.3cm}|P{1.8cm}|}
\hline
\textbf{Ref.} & \textbf{Information-Theoretic Security} & \textbf{Anti-Jamming} & \textbf{Anti-Spoofing} & \textbf{Security Research Challenges} & \textbf{Machine Learning} & \textbf{Cryptography} \\ \hline
\cite{manulis2020cyber} & \xmark & \xmark & \xmark & \cmark & \xmark & \cmark\\ \hline
\cite{Kodheli2021_comst} & \xmark & \xmark & \xmark & \xmark & \cmark & \xmark\\ \hline
\cite{Li2020_iotj} & \cmark & \xmark & \xmark & \cmark & \xmark & \xmark\\ \hline
\cite{zidan2020gnss} & \xmark & \cmark & \cmark & \cmark & \xmark & \xmark\\ \hline
\cite{morales2019_comst} & \xmark & \cmark & \cmark & \cmark & \xmark & \xmark \\ \hline
\cite{rath2020security} & \xmark & \xmark & \xmark & \cmark & \cmark & \xmark\\ \hline
\cite{junzhi2019research} & \xmark & \xmark & \cmark & \cmark & \xmark & \cmark\\ \hline
\cite{margaria_2017_ieeesignal} & \xmark & \xmark & \cmark & \xmark & \xmark & \cmark\\ \hline
\cite{saroj2016survey} & \xmark & \xmark & \cmark & \cmark & \xmark & \cmark \\ \hline
\cite{radhakrishnan2016survey} & \xmark & \xmark & \xmark & \xmark & \xmark & \xmark\\ \hline
\cite{schmidt2016survey} & \cmark & \xmark & \cmark & \cmark & \xmark & \cmark\\ \hline
\cite{hosseinidehaj2018_comst} & \cmark & \xmark & \xmark & \cmark & \xmark & \xmark \\ \hline
\textbf{This Survey} & \cmark & \cmark & \cmark & \cmark & \cmark & \cmark\\ \hline
\end{tabular}
\label{tab:related_with_features}
\end{table}
Looking at the main contributions on \ac{SATCOM} security, valuable insights are provided by, e.g. \cite{zidan2020gnss}, \cite{morales2019_comst}, \cite{junzhi2019research}, \cite{schmidt2016survey}, \cite{margaria_2017_ieeesignal}, and \cite{saroj2016survey}. However, they only preliminarily identified the \ac{GNSS} spoofing and \ac{GNSS} jamming attacks and presented the adopted solutions and mitigation techniques available from the literature.
Moreover, the study by the authors in \cite{Li2020_iotj} only summarize the schemes targeting data secrecy at the physical layer, with a focus on information-theoretic schemes only. Other surveys, such as the one in~\cite{hosseinidehaj2018_comst}, focused specifically on some research areas, such as quantum computing, missing the contextualization of such areas to other research in the SATCOM domain. 
It is also worth  mentioning the recent contribution by~\cite{guo2022_comst}, highlighting the threats affecting specific application areas where SATCOM links play a role. However, such a survey focuses on the review of the literature, rather than on the identification of the research areas where such contributions are provided.
\textcolor{black}{Moreover, although many contributions are available on secure routing, e.g.~\cite{xu2021_twc} and~\cite{xu2021_iotj}, the focus of our investigation is link-layer security in satellite communications, and secure routing is therefore out of of scope.}
As a result, we notice that the current literature is still missing a comprehensive survey, presenting and exploring all the facets of the threat surface to be considered when deploying \ac{SATCOM} systems at the link-layer, as well as related countermeasures.

\begin{figure*}
  \centering
  \includegraphics[angle=0, width=\textwidth]{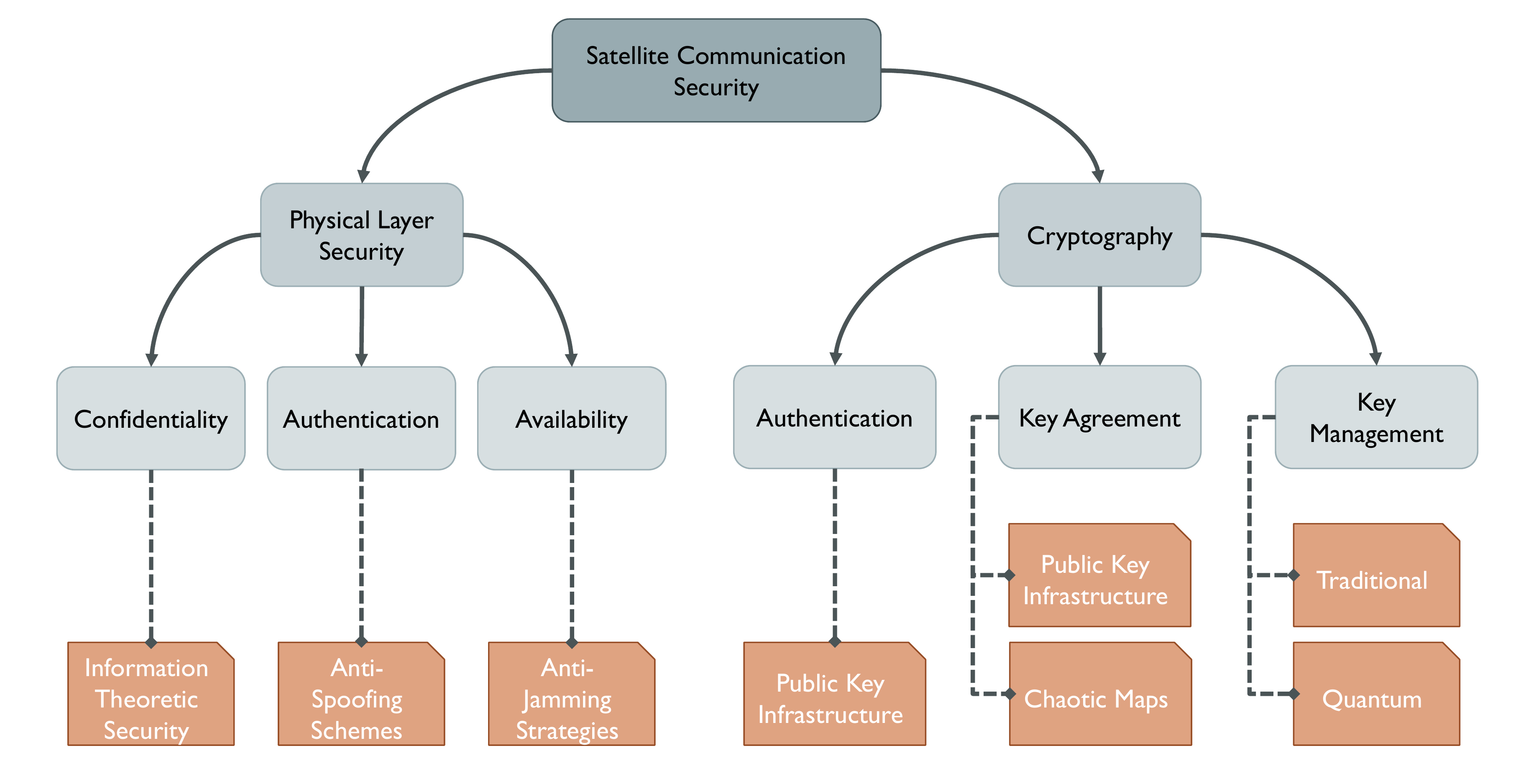}
  \caption{Taxonomy and classification of the major scientific contributions dealing with security in SATCOM. We identified two major research streams, i.e., physical-layer security and cryptography solutions. Within each stream, we extract specific research domains, where several different solutions are available.}
  \label{fig:taxonomy}
\end{figure*}

{\bf Contribution.} In this paper, we fill the afore-described gaps by providing a comprehensive survey on security threats, solutions, mitigation strategies, and research challenges faced when designing and deploying secure \ac{SATCOM} systems. 
In detail, we classify the security solutions related to the link-layer of \ac{SATCOM} systems available in the literature into two main research domains, i.e., physical-layer approaches and cryptography techniques. Next, we delve into each area, looking at the offered security services and how such schemes guarantee the desired security objectives. 
For each area, as a novel contribution, we describe the threat models, assumptions, system requirements, and operational strategy, and we cross-compare the most important proposals along their characterizing features (see Figure~\ref{fig:taxonomy} for a graphical overview). Finally, within each research domain, we also identify novel future research directions and additional research challenges.

{\bf Roadmap.} The rest of this paper is organized as follows. Section~\ref{sec:background} introduces the basics of \acp{SATCOM}; Section~\ref{sec:pls} introduces  information-theoretic security strategies and solutions to enhance data secrecy; Section~\ref{sec:cryptography} analyzes the applications of cryptography schemes in SATCOMs; Section~\ref{sec:challenge} outlines emerging research domains; and, finally, in Section~\ref{sec:conclusion} we tighten some conclusions. Refer to Table~\ref{tab:acronyms} for the acronyms list.

\begin{table}[!htbp]
    \centering
    \tiny
    \caption{Acronym List.}
    \begin{tabular}{c||l}
\textbf{Abbreviation} & \textbf{Definition}\\
\hline
AES & Advanced Encryption Standard\\
\hline
AF & Amplify and Forward\\
\hline
AGC & Automatic Gain Control\\
\hline
AI & Artificial Intelligence\\
\hline
AWGN & Additive White Gaussian Noise\\
\hline
CNR & Carrier-to-Noise Ratio\\
\hline
CPS & Cyber-Physical Systems\\
\hline
CRN & Cognitive Radio Network\\
\hline
CSI & Channel State Information\\
\hline
D2D & Device-to-Device Communications\\
\hline
DF & Decode and Forward\\
\hline
DoS & Denial of Service\\
\hline
ECC & Elliptic Curve Cryptography\\
\hline
ECDSA & Elliptic Curve Digital Signature Algorithm\\
\hline
ESA & European Space Agency\\
\hline
ESR & Ergodic Secrecy Rate\\
\hline
FDMA & Frequency Division Multiple Access\\
\hline
GG & Ground-to-Ground\\
\hline
GNSS & Global Navigation Satellite System\\
\hline
GPS & Global Positioning System\\
\hline
IoST & Internet of Space Things\\
\hline
IoT & Internet of Things\\
\hline
MIMO & Multiple-Input Multiple-Output\\
\hline
MISO & Multiple-Input Single-Output\\
\hline
MITM & Man-In-The-Middle\\
\hline
mMIMO & Massive MIMO\\
\hline
NFC & Near Field Communication\\
\hline
NIC & Network Interface Card\\
\hline
NIST & National Institute of Standards and Technology\\
\hline
NOAA & National Oceanic and Atmospheric Administration\\
\hline
NOMA & Non-Orthogonal Multiple Access\\
\hline
OFDM & Orthogonal Frequency Division Multiplexing\\
\hline
OFDMA & Orthogonal Frequency Division Multiplexing Access\\
\hline
OTP & One-Time-Pad\\
\hline
PDR & Packet Delivery Ratio\\
\hline
PS & Power Splitting\\
\hline
QoS & Quality of Service\\
\hline
RF & Radio Frequency\\
\hline
RFID & Radio Frequency IDentification\\
\hline
RSMA & Rate-Splitting Multiple Access\\
\hline
SATCOM & Satellite-based Communication\\
\hline
SDN & Software Defined Networking\\
\hline
SDR & Software Defined Radio\\
\hline
SEE & Secrecy Energy Efficiency\\
\hline
SG & Satellite-to-Ground\\
\hline
SIMO & Single-Input Multiple-Output\\
\hline
SINR & Signal to Noise plus Interference Ratio\\
\hline
SISO & Single-Input Single-Output\\
\hline
SNR & Signal-to-Noise Ratio\\
\hline
SOP & Secrecy Outage Probability\\
\hline
SRM & Secrecy Rate Maximization\\
\hline
SS & Satellite-to-Satellite\\
\hline
TDMA & Time Division Multiple Access\\
\hline
TS & Time Splitting\\
\hline
UAV & Unmanned Aerial Vehicles\\
\hline
VLC & Visible Light Communications\\
\hline
VSAT & Very Small Aperture Terminal\\
    \end{tabular}
    \label{tab:acronyms}
\end{table}

\section{Background}
\label{sec:background}

This section introduces the main notions related to \ac{SATCOM} systems used within our manuscript, including the satellites constellations, architectures, and the involved protocols. Overall, this section aims to provide the reader with the needed background on  \ac{SATCOM} technologies and their main features, that will be used in the sequel of the paper.

\subsection{Satellite Constellations}
\label{subsec:background}

The main features that distinguish satellites orbits are the shape (circular or elliptical), the altitude (Low-Earth, Medium-Earth, or Geostationary), the travel direction (clockwise or counterclockwise), and the inclination to the plane of the Earth's equator~\cite{maini2014}. The most popular of the previously cited features is the altitude: we distinguish Low Earth Orbit (LEO), Medium Earth Orbit (MEO) and Geostationary Equatorial Orbit (GEO). The respective altitude ranges from the Earth surface are $500$ to $900$ km for LEO, $5,000$ to $25,000$ km for MEO, and $36,000$ km for GEO~\cite{elbert2003_book}. The altitude is directly related to the services offered to the end users. Without loss of generality, the farther is the satellite from the Earth surface, the greater is the Earth coverage area. Indeed, the Earth Coverage for LEO satellites is quite small, for MEO is larger and for GEO is sizeable. For instance, according to the authors in~\cite{cakaj2021_fcn}, a LEO satellite located $550$~km over the Earth surface, having an elevation of $40$ degrees, can cover an area of approx. $1.05$~million km squared, with an approximate radius of $580$~km. At the same time, according to the authors in~\cite{peterson2003_satcom}, a GEO satellite can cover $40$ degrees of latitude, i.e., approximately one-third of the Earth surface. Given that the coverage range is directly related to the number of satellites to be operated, such a number decreases when the distance from the Earth increases.

\ac{SATCOM} uplink and downlink channels adopt different frequencies  to mitigate the interference on the ground and at the satellite. The band, frequency regulations, and recommendations are authorized by Federal Communications Commission (FCC) and International Telecommunications Union (ITU). For instance, according to the \ac{ESA}~\cite{esa}, SATCOMs frequency bands are standardized in the range of $1\sim40$~GHz, as reported in Table~\ref{tab:sat_frquenc_bands}.
\begin{table}[ht]
    \centering
    \caption{Satellites Frequency Bands and Applications}
    \begin{tabular}{|P{2cm}|P{1cm}|P{7.5cm}|}
        \hline
         \textbf{Satellite Frequency [GHz]} & \textbf{Band Name} & \textbf{Applications}  \\
         \hline
         $1-2$ & L & Positioning Systems, Mobile Phones, Sea/Land/Air Communications, Radio \\
         \hline
         $2-4$ & S & NASA communications with Space Shuttle and International Space Station \\
         \hline
         $4-8$ & C & Satellite TV/feed \\
         \hline
         $8-12$ & X & Military, radar (continuous-wave, pulsed, single-polarisation, dual- polarisation, synthetic aperture radar, phased arrays), weather monitoring, air traffic control, maritime vessel traffic control, defence tracking, vehicle speed detection \\
         \hline
         $12-18$ & Ku & Broadcast satellites\\
         \hline
         $26-40$ & Ka & Close-range targeting radars on military aircraft \\
         \hline
    \end{tabular}
    \label{tab:sat_frquenc_bands}
\end{table}

For instance, Inmarsat~\cite{Ilcev2018} is a provider of \ac{SATCOM} services that adopt GEO satellites to provide telephone and data services to users worldwide.
Companies like SpaceX~\cite{jeeff2019_spectrum} and Iridium~\cite{caprolu2020_commag,oligeri2020_wisec} are planning to launch in orbit thousands of LEO satellites to provide low latency, broadband internet systems, voice, and data services anywhere on Earth.

\subsection{Communication Architecture}
\label{subsec:comm_arch}

\begin{figure}[H]
  \centering
  \includegraphics[angle=0, width=\columnwidth]{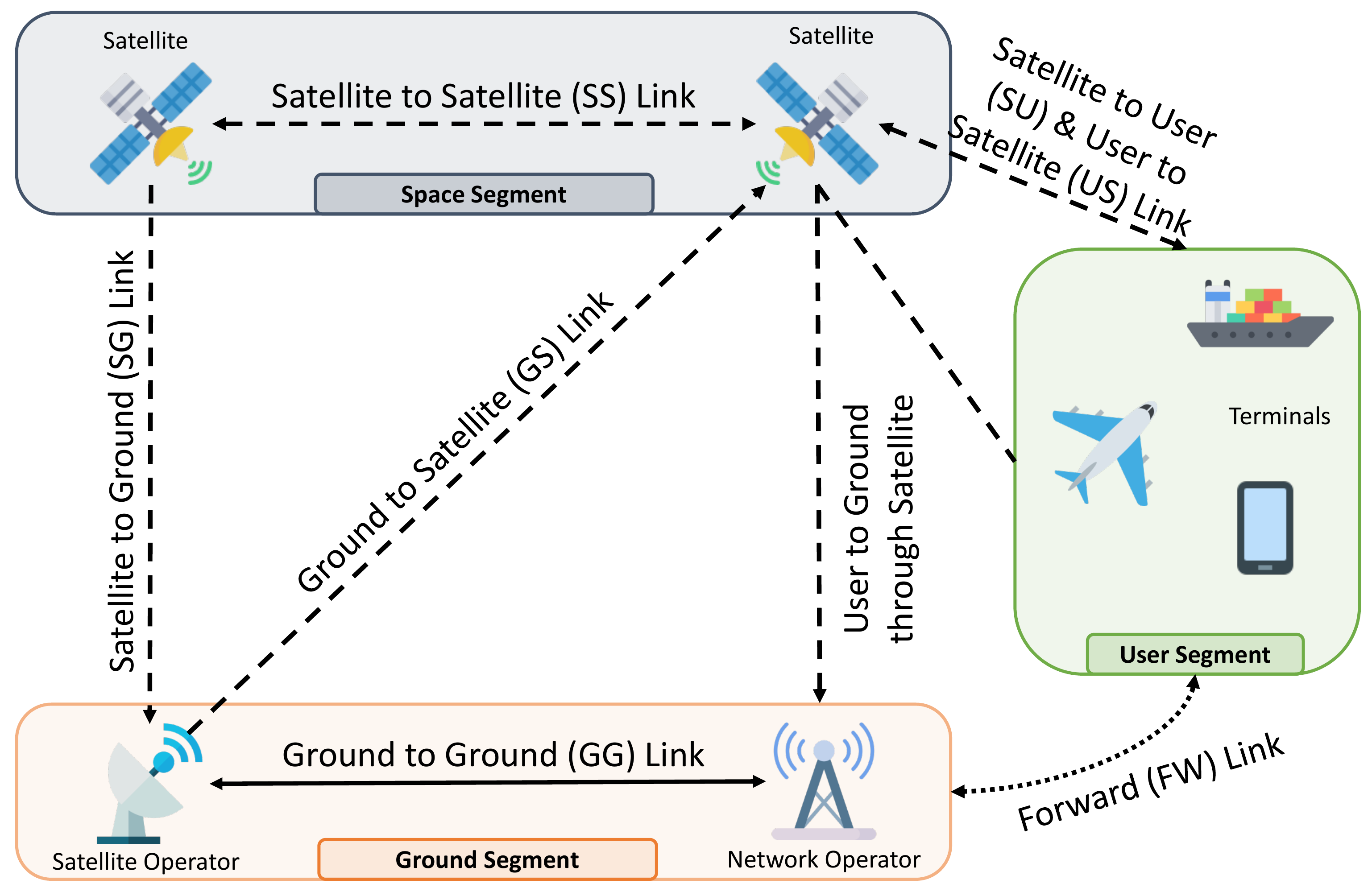}
  \caption{\ac{SATCOM} Architecture.}
  \label{fig:architecture}
\end{figure}

The reference communication architecture of a SATCOM system, as depicted in Figure~\ref{fig:architecture}, is generally characterized by: (i) a space segment including the \emph{Satellite to Satellite} (SS) and the \emph{Satellite to Ground} (SG) links; (ii) a ground segment, defined by the satellite operators (or gateways) and network operators, enabling the \emph{Ground to Satellite} (GS), \emph{Ground to Ground} (GG), \emph{Satellite to Ground} (SG), forwarding, and the \emph{Satellite to User} (SU) links; and finally, (iii) a user segment, which includes the terminals, e.g., ships, airplanes, and satellite smartphones, enabling the additional \emph{User to Ground} (UG) and the \emph{User to Satellite} (US) links.
The space segment of a SATCOM architecture is one of the three main components of a SATCOM system. This segment comprises GEO satellites to support business in navigation, data, mobile television, and radio broadcasting systems. At the same time, MEO satellites are deployed to deliver low-latency and high-bandwidth data connectivity to service providers, agencies and industries, and to support the network connectivity in the avionic/maritime domain. LEO satellite constellations are also adopted for several applications such as imaging, and low-bandwidth telecommunications and broadband internet. Each of these satellites is placed in orbit by a launch vehicle. The space segment also includes military and defense communication systems, as well as commercial \ac{SATCOM} transponders and payloads. Note that the afore-mentioned communication links involving satellites all use frequencies in the L-band, in the range $[ 1 -- 2]$~GHz.

The ground segment can help to establish the communication between the satellites and all the terminals defined in the user segment. It comprises dedicated Gateway stations, namely Satellite Operator, infrastructures for control, Network Operator such as the Network Control Centre (NCC) and the Network Management Centre (NMC) supporting the satellite access requests from users. 

The user segment includes the user terminals, such as satellite mobile phones, ships, and airplane, to name a few. These devices can communicate with satellites by leveraging the link between the ground segment and the user segment, such as the forward link~\cite{yuma2018_ieice}, while their communication with the gateways can take place over any communication technology. The forward link consists of both an uplink (base station to satellite) and a downlink (satellite to mobile user). Conversely, constellations like Iridium, Globalstar, Thuraya and Inmarsat allow a direct connection of the user handsets to the satellites, using the \emph{User to Satellite} (US) link that typically uses frequencies in the L-band.

While the above discussion  covers the traditional SATCOM architecture, many variations can be found. In this context, it is worth noting that the 3GPP issued several standards over the last few years, with the objective to define the general architecture of non-terrestrial satellite networks in the context of 5G networks. Such standards are meant for several types of non-terrestrial communications, including GNSS, High-Altitude Platform Systems (HAPS), and air-to-ground communications---illustrating use-cases, scenarios, channels to be used, and modulation formats, to name a few~\cite{lin2021_commag}. 

Finally, from the security perspective, note that attacks can be launched in any of the identified segments. Thus, the ground segment should be appropriately secured, as well as any communication originating from the satellite should be protected, independently from its target destination (ground or space). More details on the specific threats will be provided in the following sections.

\section{Physical Layer Security Schemes for SATCOM}
\label{sec:pls}

In this section, we introduce, review, and classify approaches investigating security issues of \ac{SATCOM} technologies at the physical layer. Specifically, Section~\ref{sec:information_theoretic} introduces information-theoretic security approaches for physical layer confidentiality of communications, Section~\ref{sec:gnss_spoofing} focuses on \ac{GNSS} anti-spoofing techniques, while Section~\ref{sec:jamming} includes considerations on anti-jamming solutions of \ac{SATCOM}. To conclude our critical discussion, Section~\ref{sec:pls_lessons} summarizes the main lessons learned while Section~\ref{sec:pls_future} highlights future research directions in the area of physical-layer security.

\subsection{Information Theoretic Security}
\label{sec:information_theoretic}

\acp{SATCOM} are particularly prone to eavesdropping due to the broadcast nature of the wireless medium and the very large coverage area. Usually, the confidentiality of \ac{SATCOM} communications is provided via traditional cryptographic protocols such as \ac{AES}, working at the MAC-layer or above. However, legacy satellites deployments often use old and proprietary customized versions of AES, frequently found later to be insecure. As a result, motivated adversaries featuring powerful capabilities and tools can easily collect a consistent amount of encrypted data and possibly compromising communications confidentiality. Moreover, many satellites deployments were set up several years ago, when wireless security was not conceived as a requirement. Indeed, attacks on \ac{SATCOM} channels was conceived by the operators as hard to achieve, and overall, security was thought as a slow-down factor rather than an enabler. 
Thus, many satellites do not implement any security protection, and updating them today would require high costs~\cite{aptsky}.

Taking into account the  above exposed issues, in the last years many contributions proposed to provide confidentiality to \ac{SATCOM} scenarios by applying information-theoretic security schemes. The rationale supporting such schemes is the following:
information-theoretic security approaches leverage the inherent randomness
and noise of \ac{SATCOM} communication channels to minimize the amount of information that can be extracted at the PHY layer by an unauthorized receiver~\cite{mukherjee2014_comst}. This is typically achieved by guaranteeing that the quality of the channel, expressed in terms of \ac{SNR}, exceeds a given bound only at the authorized receiver's location while remaining below the set threshold 
in other locations, so that an eavesdropper cannot decode the received message. 
Moreover, such schemes typically do not assume any constraints for the eavesdropper in terms of processing capabilities or network parameter knowledge, and the resulting security features can be quantified analytically via dedicated channel-level metrics. The afore-mentioned security features are achieved without resorting to  cryptography materials (e.g. shared key, certificates) or  crypto related computations, hence resulting  in a very efficient solution (from the computational,  storage, and bandwidth point of view).

We present a comprehensive classification of the scientific contributions that achieve confidentiality via information-theoretic schemes in SATCOM scenarios in Table~\ref{tab:information_theoretic_security}. 
\begin{table}[htbp]
	\centering
	\scriptsize
	\caption{Comparison of scientific contributions adopting information-theoretic approaches for \ac{SATCOM} confidentiality. }
		\begin{tabular}{|P{0.45cm}|P{0.5cm}|P{1.8cm}|P{1.3cm}|P{1.5cm}|P{2.3cm}|P{1.6cm}|}
		\hline
        \textbf{Ref.} & \textbf{Link} & \textbf{CSI} & \textbf{Adversary} & \textbf{Adversary Antennas} & \textbf{Adversary Antenna Type} & \textbf{Performance Metrics} \\
        \hline
        \cite{an2016_jsac} & SG & Imperfect, Statistical & External & Single & Omni-Directional & SOP, SR \\
        \hline
        \cite{lei_2011_tifs} & SG & \cmark & Internal, External & Single & Omni-Directional & SR \\
        \hline
        \cite{zheng_2012_twc} & SG & \cmark & Internal & Multiple & Omni-Directional & SR \\
        \hline
        \cite{xu_2019_ieeecomlet} & SG, GS & \cmark, Imperfect & External & Single & Omni-Directional & Secrecy Capacity (SC) \\
        \hline
        \cite{luo_2017_access} & SG & \xmark & External & Dual Polarized Antenna & Omni-Directional & Polarization Filtering \\
        \hline
        \cite{lu_2019_taes} & SG & Imperfect & Internal & Single & Omni-Directional & Sum SR \\
        \hline
        \cite{xu_2019_ietcomm} & SG, GS & \cmark, Statistical & Internal & Single & Omni-Directional & SC \\
        \hline
        \cite{lin_2018_jsac} & SG & \cmark & External & Single & Omni-Directional & SR \\
        \hline
        \cite{zhang_2017_access} & SG & \cmark & External & Unipolar Parabolic Antennas & Omni-Directional & SR, SC \\
        \hline
        \cite{lin_2018_jsac2} & SG & \cmark, Imperfect & External & Single & Omni-Directional & SR \\
        \hline
        \cite{kalantari_2015_tifs} &  SG, GS & \cmark & External & Single & Omni-Directional & Sum SR \\
        \hline
        \cite{huang_2017_ietcomm} & SG & \cmark & External & Single & Omni-Directional & SR \\
        \hline
        \cite{yan_2016_wocc} & SG & \cmark & External & Single & Omni-Directional & SR\\
        \hline
        \cite{liu_2018_cse} & SG & \xmark & External & Single & Omni-Directional & SC, SNR, BER \\
        \hline
        \cite{an_2016_wocc2} & SG & \cmark & External & Single & Omni-Directional & SC, SOP \\
        \hline
        \cite{li_2018_ieeewcl} & SG & Imperfect & External & Multiple & Omni-Directional & SR \\
        \hline
        \cite{li_2018_tvt} & SG & Imperfect & External & Single & Omni-Directional & SR \\
        \hline
        \cite{guo_2019_access} & SG & \cmark & External & Single & Omni-Directional & SOP, SC \\
        \hline
        \cite{xu_2019_access} & SG, GS & \cmark, Imperfect & External & Single & Omni-Directional & SC, SOP \\
        \hline
        \cite{knopp2021} & SG & \cmark & External & Single & Omni-Directional & SC \\
        \hline
	    \end{tabular}
	\label{tab:information_theoretic_security}%
\end{table}
In the following, we summarize the most important features identified throughout our analysis.

{\bf Performance Metrics.} Many performance metrics can be used to evaluate the effectiveness of a scheme proposed to ensure confidentiality at the PHY-layer via information-theoretic schemes, including, e.g., the secrecy capacity/secrecy rate, average secrecy rate, and secrecy outage probability, to name a few. We hereby present the definition of the most important ones, leading to the definition of all the others.

We assume that a malicious user $E$ is interested in eavesdropping the traffic exchanged between two legitimate network entities and correctly decoding the information. The \emph{Secrecy Rate} of the communication link is defined as the difference between the capacity of the eavesdropper $E$ and the capacity of the legitimate user~\cite{Aliberti2016}, where the \textit{capacity} is the maximum transmission rate at which an eavesdropper is unable to decode any information. The \emph{secrecy rate} of the legitimate source-to-destination communication link at the physical layer is defined as in the following Eq.~\ref{eq:sec_rate}:
\begin{equation}
    \label{eq:sec_rate}
    S = C_L - C_E,
\end{equation}
where $C_L$ is the capacity of the legitimate channel and $C_E$ is the capacity of the channel source-eavesdropper. 

Another important metric is the \emph{\ac{SOP}}, defined as the probability that the instantaneous secrecy capacity drops below a specific threshold value, representing a target secrecy rate~\cite{barros2006secrecy}. The other metrics listed in Table~\ref{tab:information_theoretic_security} can be derived from the ones previously introduced. Additional details can be found in surveys dedicated to the topic, such as \cite{mukherjee2014_comst}.

{\bf Link.} The largest part of the analyzed works considered the Satellite-to-Ground (SG) link, applying information-theoretic approaches to secure the communications from the satellite to the ground receivers. To the best of our knowledge, only four works considered the Ground-to-Satellite link, i.e., \cite{xu_2019_ieeecomlet}, \cite{xu_2019_ietcomm}, \cite{kalantari_2015_tifs}, and \cite{xu_2019_access}, while other links are never considered.

{\bf CSI Availability.} One of the prominent features allowing to compare the contributions on physical-layer security is the amount of information known to the network about the attacker. Looking at the adversary model, the most restrictive assumption is the complete unavailability of information about the channel experienced by the eavesdropper (in technical terms, this is the \ac{CSI}). 
The vast majority of the works assume the perfect knowledge of the channel quality at the eavesdropper side. This is the most secure approach to analyze the problem from an information-theoretic perspective, as it allows to maximize in  analytical terms the difference between the legitimate channel and the (potentially) eavesdropped one, i.e. the secrecy rate. In practice, the cited objective is guaranteed by either reducing the probability of correct signal decoding by the adversary, or maximizing the secrecy rate of the main communication links.
Few other approaches, instead, assume to know either partially (Imperfect) or completely the channel experienced by potential eavesdroppers. Such adversary model is often referred to as an \emph{active eavesdropper}, as it is a legitimate network node, interacting with the network at times (so, its \ac{CSI} parameter is known), but also equipped with eavesdropping capabilities on other communications. A few contributions, such as~\cite{an2016_jsac, xu_2019_ieeecomlet, lu_2019_taes, lin_2018_jsac2, li_2018_ieeewcl, li_2018_tvt, xu_2019_access} evaluated the impact of the aforementioned assumption on the related security metrics.

\textbf{Adversary.} In line with related work, we consider two types of adversaries: the internal and the external one. In particular, we define an \emph{internal} adversary as an attacker playing the role of a legitimate network entity, actively participating in the network activities by transmitting and receiving information, such as an \emph{active eavesdropper}~\cite{tedeschi2020_cst}. We define an \emph{external} adversary as an attacker who is not part of a legitimate SATCOM, also striving to remain not detected--- \emph{hidden}. This latter category also comprises the
{\em external eavesdropper}, needing a simple receiving antenna tuned on the same frequency of the legitimate communication channel to receive packets successfully.

\textbf{Adversarial Receiving Antennas.} The simplest adversary model, considered by most of the analyzed contributions, is a single eavesdropper, not sharing any information with other receivers~\cite{an2016_jsac, lei_2011_tifs, xu_2019_ieeecomlet, lu_2019_taes, xu_2019_ietcomm, lin_2018_jsac, lin_2018_jsac2, kalantari_2015_tifs, huang_2017_ietcomm, yan_2016_wocc, liu_2018_cse, an_2016_wocc2, li_2018_tvt, guo_2019_access, xu_2019_access}. We remark that this is the easiest to analyze from the security perspective, as the previously-mentioned performance metrics only have to consider a single adversary. Just a few works, i.e., \cite{li_2018_ieeewcl} and \cite{zhang_2017_access}, considered multiple antennas, even if no collusion between them is considered, thus reducing the adversary model to the same of multi single-antenna adversaries.

\textbf{Adversarial Antenna Type.} All the analyzed contributions consider adversaries equipped with omnidirectional antennas, i.e., antennas capturing the information independently from the source location and (partially) radio environment. 
While such a model could appear the strongest, it does not consider a realistic \ac{SATCOM} scenario, where obstacles at the ground could alter the profile of the received signal, as well as the previously-mentioned performance metrics.

\subsection{Anti-Spoofing Schemes}
\label{sec:gnss_spoofing}

Without loss of generality, \emph{spoofing} refers to disguising a communication from an unknown source as being from a known, trusted source~\cite{schmidt2016survey}. 

Like any other wireless communication technology, \acp{SATCOM} are in principle vulnerable to spoofing attacks. The issue is even more cogent because of the presence of legacy deployments, as previously described in Section~\ref{sec:information_theoretic}. Today, several satellite systems transmit either unauthenticated messages, or authenticated at the application layer, via either symmetric key (implicit authentication) or public key solutions. A few examples include \ac{GNSS} technologies such as \ac{GPS}, Beidou, Glonass, and Galileo, and weather satellites such as \ac{NOAA} and Meteor~\cite{sciancalepore2018_cns}.

Given the considerable threat surface, the design of anti-spoofing techniques for \ac{SATCOM} scenarios has been largely focused on \ac{GNSS} technologies due to their widespread use  and increasing importance in the modern connected society~\cite{wu2020_access}. Today, thanks to the sensitive advances in the design of \acp{SDR}, performing \ac{GNSS} spoofing attacks is surprisingly easy. An attacker needs an \ac{SDR} and an omnidirectional transmitting antenna; by downloading freely-available tools such as \texttt{gps-sdr-sim}~\cite{gps_sdr_sim}, the attacker just needs to run a script to emulate a complete satellite constellation and move the target wherever in the world, also for a significant period of time~\cite{oligeri2019_wisec}. Considering that \ac{GNSS} satellites are also used for time synchronization in several IoT deployments, detecting spoofing attacks with lightweight and effective techniques is of paramount importance. 

Table~\ref{tab:spoofing} summarizes the most important contributions in the topic of \ac{GNSS} anti-spoofing and cross-compares them across reference system features.
\begin{table*}[htbp]
    \scriptsize
	\centering
	\caption{Comparison of \ac{GPS}/\ac{GNSS} Spoofing Detection Methods and Related System Requirements.}
		\begin{tabular}{|P{.5cm}|P{5.4cm}|P{1.4cm}|P{1.7cm}|P{1.7cm}|P{1.4cm}|P{1.9cm}|}
		\hline
        \textbf{Ref.} & \textbf{\ac{GNSS} Spoofing Detection Means} & \textbf{No need of  Multiple Antennas} & \textbf{No PHY-layer information Required} & \textbf{No need of Ad-Hoc Network Infrastructure} & \textbf{No need of Dedicated Hardware}\\
        \hline
        \cite{humphreys_2013_taes} & Statistics Approach & \cmark & \cmark & \cmark & \cmark\\
        \hline
        \cite{psiaki_2013_iongnssm} & HF Antenna Motion \& Carrier-Phase & \cmark & \xmark & \cmark & \xmark\\
        \hline
        \cite{ieee_2013_taes} & Phase-Only Analysis of Variance & \xmark & \xmark & \cmark & \xmark\\
        \hline
        \cite{wesson_2013_gcsip} & Symmetric Difference Autocorrelation Distortion Monitor and a Total in-band Power Monitor & \cmark & \xmark & \cmark & \xmark\\
        \hline
        \cite{sciancalepore2018_cns} & Meteor Burst Communications & \xmark & \xmark & \xmark & \cmark\\
        \hline
        \cite{oligeri2019_wisec} & Cellular Network & \cmark & \cmark & \cmark & \cmark\\
        \hline
        \cite{heng2014gps} & Cross-Check Receivers & \xmark & \xmark & \xmark & \xmark\\
        \hline
        \cite{yu2014short} & Multilateration Phasor Measurement Units in Smart Grids & \xmark & \cmark & \xmark & \xmark\\
        \hline
        \cite{heng2013cooperative} & Cross-Correlation and Cooperative Authentication & \cmark & \cmark & \xmark & \cmark\\
        \hline
        \cite{psiaki2014gnss, stenberg2020gnss} & Carrier-Phase Measurements & \xmark & \xmark & \cmark & \xmark\\
        \hline
        \cite{o2013real} & Code Signals Correlation & \cmark & \cmark & \cmark & \xmark\\
        \hline
        \cite{hu2018gnss} & Total Signals Energy Measurement & \cmark & \xmark & \cmark & \xmark\\
        \hline
        \cite{bhamidipati2018gps} & Time Authentication & \cmark & \xmark & \xmark & \xmark\\
        \hline
        \cite{mina2018detecting} & Multi-Receiver Hybrid Communication Network for Power Grid Timing Verification & \cmark & \xmark & \xmark & \xmark\\
        \hline
        \cite{hu2018gnss2} & Fraction Parts of Double-difference Carrier Phases & \xmark & \xmark & \cmark & \xmark\\
        \hline
        \cite{formaggio2018authentication, formaggio2019authentication} & Channel Gain / Estimation Noise & \cmark & \cmark & \cmark & \cmark\\
        \hline
        \cite{schmidt2020gps} & Least Absolute Shrinkage and Selection Operator & \cmark & \xmark & \cmark & \cmark\\
        \hline
        \cite{anderson2017chips} & Chips-Message Robust Authentication & \cmark & \xmark & \cmark & \xmark\\
        \hline
        \cite{tohidi2020effective} & Neural Network & \cmark & \cmark & \cmark & \cmark\\
        \hline
        \cite{gross2018maximum, wang2017gnss, wesson2017gnss} & Maximum-Likelihood & \cmark & \xmark & \cmark & \xmark\\
        \hline
        \cite{jiang2018satellite} & K-mean clustering & \cmark & \xmark & \cmark & \cmark\\
        \hline
        \cite{zou2016detection} & Control Theory (IMU sensor) in UAVs & \cmark & \cmark & \cmark & \cmark\\
        \hline
        \cite{axell2015gnss} & Cooperative Receivers Positions & \cmark & \cmark & \xmark & \cmark\\
        \hline
        \cite{caparra2016autonomous} & Semi-Codeless Receiver & \cmark & \xmark & \cmark & \xmark\\
        \hline
        \cite{singh2020mitigating} & Genetic Algorithm, Shortest Path and Pattern Matching & \cmark & \cmark & \cmark & \cmark\\
        \hline
        \cite{semanjski2020gnss} & Supervised Machine Learning & \cmark & \xmark & \cmark & \cmark\\
        \hline
        \cite{oligeri2020_wisec} & IRIDIUM Ring Alert & \cmark & \cmark & \cmark & \cmark \\
        \hline
	    \end{tabular}
	\label{tab:spoofing}%
\end{table*}

{\bf \ac{GNSS} Spoofing Detection Means.} A large variety of means have been used to detect \ac{GNSS} spoofing attacks. Approaches use either PHY-layer information (\cite{psiaki_2013_iongnssm,ieee_2013_taes,wesson_2013_gcsip,sciancalepore2018_cns,heng2014gps,psiaki2014gnss,stenberg2020gnss,hu2018gnss,bhamidipati2018gps,mina2018detecting,hu2018gnss2,schmidt2020gps,anderson2017chips,gross2018maximum, wang2017gnss, wesson2017gnss,jiang2018satellite,caparra2016autonomous,semanjski2020gnss}), or additional communication technologies (\cite{humphreys_2013_taes,oligeri2019_wisec,yu2014short,heng2013cooperative,o2013real,oligeri2020_wisec}), or techniques based on \ac{ML} over heterogeneous data (\cite{tohidi2020effective,gross2018maximum, wang2017gnss, wesson2017gnss,jiang2018satellite,singh2020mitigating,semanjski2020gnss}). 
All these techniques share the basic consideration that the bitstrings in \ac{GNSS} signals cannot be modified to be more secure before being transmitted. Indeed, such modifications woul require temporarily stopping the operation of the \ac{GNSS} satellite, causing significant and unmanageable costs and effort.

{\bf Usage of Multiple Receiving Antennas.} Simplest \ac{GNSS} spoofing attacks assume that the adversary transmits a fake \ac{GNSS} signal using only a single antenna. Signal cross-correlation detection methods identify such attacks by discriminating the injected signal from the expected one, as it is transmitted using different devices than the ones the target satellite is equipped with. Comparing the correlation of the received signals using more than one antenna at the receiver side will result in a significant difference in signal characteristics, such as the carrier phase and amplitude. This is just an example of a technique employing multiple receiving antennas to detect \ac{GNSS} spoofing, and other examples include the contributions by the authors in~\cite{ieee_2013_taes, sciancalepore2018_cns, heng2014gps, yu2014short, psiaki2014gnss, stenberg2020gnss, hu2018gnss2}. Although being relatively cheap to deploy, such solutions usually require the deployment of multiple antennas and their connection to a single system, i.e., the hardware modification of the receiving device. Often, such operations might be too expensive or impractical to be applied due to application-specific limitations.

{\bf Usage of PHY-layer information.} \ac{GNSS} messages transmitted by satellites go through different phases of signal processing before being converted into a digital form at the receiver side. The anomalies in the characteristics of signals during any of these processing phases could be used to detect counterfeit \ac{GNSS} signals~\cite{falco2019algorithm}. Anomalies can be found in signals features such as received power, \ac{CNR}, quality, correlation, \ac{AGC}, clock bias, and angle of arrival, to name a few. Such features are used in many approaches, e.g.,  \cite{psiaki_2013_iongnssm, ieee_2013_taes, wesson_2013_gcsip, bhamidipati2018gps}, to detect \ac{GNSS} spoofing attacks. On the one hand, such techniques could be compelling and reliable. On the other hand, they require the receiving devices to access such information, which is not always possible. Indeed, many modern chipsets do not provide PHY-layer information to the devices they are connected to or integrated into, preventing the application of such approaches.

{\bf Usage of Ad-Hoc Network Infrastructures.} Other contributions leverage additional network infrastructures, set up ad-hoc for \ac{GNSS} spoofing detection. This is the case of approaches using dedicated sensors deployments (\cite{sciancalepore2018_cns,heng2014gps,yu2014short,heng2013cooperative,bhamidipati2018gps,mina2018detecting,axell2015gnss}). Other approaches, such as~\cite{oligeri2019_wisec} and~\cite{oligeri2020_wisec}, use opportunistic signals gathered by other communication infrastructures, such as the cellular network and the IRIDIUM constellation. While the first set of approaches require a dedicated setup, that is not always possible, the security of the second class of approaches mostly depends on the adversary model and on its capability to spoof also the additional wireless signals.

{\bf Usage of Dedicated Hardware.} Many solutions have been proposed in the last years, recurring to dedicated hardware to detect \ac{GNSS} spoofing attacks. Such approaches leverage either specific information available from the radio channel (\cite{psiaki_2013_iongnssm,ieee_2013_taes,wesson_2013_gcsip,heng2014gps,psiaki2014gnss, stenberg2020gnss,hu2018gnss,bhamidipati2018gps,mina2018detecting,hu2018gnss2,anderson2017chips,gross2018maximum, wang2017gnss, wesson2017gnss,caparra2016autonomous}), or specific type of arrays of antennas, or the use of dedicated sensors providing inertial measurements~\cite{zou2016detection}. Similarly to previous approaches using multiple antennas, such techniques might be very efficient. However, they require the adoption of compatible hardware, which sometimes does not represent a viable solution.

\subsection{Anti-Jamming Strategies}
\label{sec:jamming}

In this section, we discuss the most important contributions dealing with \emph{anti-jamming} methods in \ac{SATCOM} scenarios. Without loss of generality, \emph{jamming} is defined as the injection of intentional interference into the wireless channel in a way to disrupt the operations of a legitimate communication channel~\cite{vadlamani2016}. Several classes of jammers have been proposed in the scientific literature throughout the last years~\cite{zhang_tsp}. With reference to the portion of time where they are active, jammers can be constant, alternate,  proactive (if they choose a channel in advance, and jam it)~\cite{wang2018adaptive}, or reactive (if they jam a specific channel only when RF activity is detected on that channel). Considering   the number of frequencies jammed at the same time, jammers can be spot (a single jammed frequency)~\cite{morehouse2021incremental}, sweep (multiple frequencies, at different times)~\cite{borio2016}, or barrage (multiple frequencies at the same time)~\cite{topal2020}. Finally, based on the type of signal injected to cause interference, we can have noise-jammers  (if noise of a different type, e.g., \ac{AWGN} is injected on the channel), or deceptive-jammers (if they inject a signal similar to the legitimate ones).
The effectiveness of a jammer strictly depends on the communication parameters~\cite{dempster_2016_ieee} set by both the transmitter and the receiver~\cite{borio_2016_ieee}. Overall, to maximize the performance of the jammer, the adversary should carefully analyze the radio link and then design and deploy the appropriate type of jammer.

Jamming is particularly relevant in the context of SATCOM links, and several contributions highlight the inefficacy of currently-deployed communication schemes, e.g., \ac{CDMA}, when powerful jammers target the communication links~\cite{jung2018_taes}. Table~\ref{tab:jamming} reviews the most important contributions providing anti-jamming techniques in the \ac{SATCOM} context and cross-compares them across reference features.
\begin{table*}[htbp]
	\centering
	\scriptsize
	\caption{Comparison of anti-jamming techniques and related system requirements.}
		\begin{tabular}{|P{0.5cm}|P{0.5cm}|P{1.5cm}|P{1.1cm}|P{2.3cm}|P{2.4cm}|P{1.5cm}|P{1.6cm}|}
		\hline
        \textbf{Ref.} & \textbf{Link} & \textbf{Technology} & \textbf{Jammers} & \textbf{Jammer Type} &  \textbf{Technique} & \textbf{No Dedicated Hardware} & \textbf{Assessment}\\
        \hline
        \cite{tamazin2020robust} & SG & \ac{GPS} & Single & Sweep, Spot & Fast Orthogonal Search & \xmark & Simulations \\
        \hline
        \cite{bavzec2016gps} & SG & \ac{GPS} & Single & AWGN, Spot & Signals Cross Correlation & \xmark & Experimental \\
        \hline
        \cite{purwar2016gps} & SG & \ac{GPS} & Single & Constant, Spot, Barrage & Turbo Codes & \cmark & Simulations\\
        \hline
        \cite{gao2017gnss} & SG & \ac{GNSS} & Single & Deceptive, AWGN & ML Classification & \cmark & Simulations \\
        \hline
        \cite{glomsvoll2017gnss} & SG & GLONASS, \ac{GPS} & Single & Sweep, Spot & Dual Frequencies Correlation & \cmark & Experimental  \\
        \hline
        \cite{gao_2016_ieee} & SG & \ac{GNSS} & Single & Generic & Various Filtering & \xmark & Analysis  \\
        \hline
        \cite{lichtman_2016_ijscn} & GS, SG & $40$~GHz UL, $20$~GHz DL & Single & Reactive & Geometric Jamming Constraints & \xmark & Simulations\\
        \hline
        \cite{borio_2015_iclgnss} & SG & \ac{GNSS} & Single & Sweep & Sum-of-Squares, Correlation & \xmark & Experimental \\
        \hline
        \cite{shi_2016_cns} & SG & \ac{SATCOM} & Multiple & Reactive & Dynamic Spectrum Access & \xmark & Simulations \\
        \hline
        \cite{borio_2016_enc} & SG & \ac{GNSS} & Single & Constant & Pulse Blanking & \xmark & Experimental \\
        \hline
        \cite{wang_2017_milcom} & GS & \ac{SATCOM} & Single & Constant, AWGN & Game Theory & \cmark & Simulations \\
        \hline
        \cite{lang_2017_plos} & SG & \ac{GPS} & Multiple & Constant, Barrage & Multi-objective optimization & \xmark & Experiments \\
        \hline
        \cite{wu_2017_isspit} & SG & \ac{SATCOM} & Single & Constant, AWGN & Convolutional Neural Network & \cmark & Simulations \\
        \hline
        \cite{sun_2015_cisp} & SG & \ac{SATCOM} & Single & Generic & Polarization Diversity & \xmark & Simulations \\
        \hline
        \cite{wang_2017_iaces} & SG & Beidou & Single & Generic & Spatial-Time Polarization & \xmark & Simulations \\
        \hline
        \cite{dong_2017_icwcsp} & SG & \ac{GNSS} & Multiple & Constant, AWGN & Cross Spectral Self-Coherence Restoral algorithm & \xmark & Simulations\\
        \hline
        \cite{wang_2019_taes} & SG & \ac{GNSS} & Multiple & Sweep & Adaptive-Partitioned Subspace Projection & \xmark & Simulations \\
        \hline
        \cite{hannon_2016_milcom} & SG & SATCOM & Single & Constant, Barrage & Frequency Hopping & \cmark & Simulations \\
        \hline
        \cite{winter_2016_milcom} & GS, GG & \ac{SATCOM} & Multiple & Constant & Maximum Ratio Combining & \cmark & Simulations  \\
        \hline
        \cite{lubbers_2015_iainwc} & SG & \ac{GPS} & Single & Constant, Barrage & Distance Theoretical Model & \cmark & Experimental \\
        \hline
        \cite{chien_2017_tfeccs} & SG & \ac{GNSS} & Multiple & Continuous Wave & Wavelet Transform & \xmark & Simulations, Experiments \\
        \hline
	    \end{tabular}
	\label{tab:jamming}%
\end{table*}

In the following, we summarize the most important considerations emerging from our analysis.

{\bf Link.} Similarly to previous physical-layer solutions, also anti-jamming strategies mostly considered the Satellite-to-Ground communication link, focusing on increasing the availability of satellite services on the ground. A few works, such as \cite{lichtman_2016_ijscn}, \cite{wang_2017_milcom}, and \cite{winter_2016_milcom}, considered also the Ground-to-Satellite link, while only one work, i.e, \cite{winter_2016_milcom}, discussed anti-jamming solutions for the Ground-to-Ground link. The Satellite-to-Satellite link is never considered because of the actual hardness  of the jamming at high distances, though further studies on this segment would be valuable..

{\bf Technology.} While some works focused on a generic \ac{SATCOM} technology, most were more specific, and analyzed the jamming issue in \ac{GNSS} (\cite{gao2017gnss,gao_2016_ieee,borio_2015_iclgnss,borio_2016_enc,dong_2017_icwcsp,wang_2019_taes,chien_2017_tfeccs}) and Military \ac{SATCOM} constellations. Others were even more focused, proposing anti-jamming schemes tailored to the specific \ac{GNSS} technology, such as the Chinese Beidou (\cite{wang_2017_iaces}), the Russian Glonass (\cite{glomsvoll2017gnss}), and the US \ac{GPS} (\cite{tamazin2020robust,bavzec2016gps,purwar2016gps,lang_2017_plos,lubbers_2015_iainwc}).

{\bf Number of Jammers.} Most of the analyzed works considered single jamming devices that are usually easier to detect and isolate from a \ac{SATCOM} link due to its vast coverage range. Other recent contributions, such \cite{lichtman_2016_ijscn, shi_2016_cns, wang_2017_milcom, wu_2017_isspit, sun_2015_cisp, winter_2016_milcom} introduced particular techniques to defend the commercial and civilian SATCOMs when one or more jammers are deployed in the scenario, thus being more effective in real deployments.  Indeed, the deployment of multiple jammers is an essential problem in tactical and military scenarios, where the adversary is so powerful to make the adoption of the current solutions challenging or impractical. 

{\bf Type of Jammer.} Characterizing and profiling the type of jamming affecting the communication link is the first step towards the deployment of an effective anti-jamming solution. Based on such considerations, the authors in \cite{shi_2016_cns,lang_2017_plos,dong_2017_icwcsp,wang_2019_taes,winter_2016_milcom,chien_2017_tfeccs} proposed anti-jamming algorithms able to characterize the jamming signals emitted from multiple jammers and still guarantee the communication quality.
In this context, it is worth mentioning the work by the authors in \cite{hannon_2016_milcom}, mitigating jamming in \ac{SATCOM} by proposing a cost-effective solution able to thwart jamming through an efficient jamming-dependent adaptive frequency hopping pattern.

{\bf Anti-Jamming Technique.} Many scientific contributions used physical layer parameters to estimate and guarantee the availability of downlink and uplink SATCOMs under jamming. For instance, to face malicious jamming attacks, the contributions in~\cite{tamazin2020robust, bavzec2016gps, purwar2016gps, glomsvoll2017gnss, lang_2017_plos, lubbers_2015_iainwc} recommended the use of well-known techniques such as fast orthogonal search, signals cross-correlation, turbo codes, and distance theoretical models for the \ac{GPS} signals.
Alternatively, the authors in \cite{glomsvoll2017gnss} demonstrate that the single-frequency multi-constellation receivers offer better jamming resilience than multi-frequency (L1 + L2) \ac{GPS} receivers and that the GLONASS constellation demonstrated a better resilience than \ac{GPS}. Indeed, they propose a multi-constellation solution that adopts \ac{GPS} and GLONASS receivers for maritime applications.

\textbf{No need of Dedicated Hardware.} It is worth noticing that contributions such as \cite{purwar2016gps, gao2017gnss, wang_2017_milcom, wu_2017_isspit, hannon_2016_milcom, winter_2016_milcom, lubbers_2015_iainwc} do not require any dedicated hardware to deploy the provided solution in a real environment. Such solutions can be implemented via simple software updates---a cheap and convenient feature. Conversely, the remaining solutions require to intervene on the hardware, and their deployment depends on the opportunity, cost, and convenience of such a modification.

{\bf Assessment Methodology.} Finally, we notice that most of the analyzed approaches were evaluated using simulations. Although simulations provide useful details into the performance of the disclosed approaches, they often miss some elements of the actual deployment, hard to be modeled and controlled into a computer-based environment. Taking into account such considerations, approaches such as~\cite{bavzec2016gps,glomsvoll2017gnss,borio_2015_iclgnss,borio_2016_enc,lubbers_2015_iainwc,chien_2017_tfeccs} worked on real deployments, showing the effectiveness of their solutions through via practical experiments or real-world data.

\subsection{Lessons Learned}
\label{sec:pls_lessons}
In the following, we summarize the main lessons learned from the investigation and cross-comparison of the approaches working on improving the security of \ac{SATCOM} deployments via physical-layer solutions.

{\bf No Satellites Hardware Update.} All the analyzed approaches do not propose the modification of the transmitted signals or the transmitting chain. Indeed, modifying a satellite is assumed to be too expensive to be performed, both from the financial and the operational perspective. Thus, assuming that the authenticity/availability received signal cannot be fully guaranteed, the studied proposals come up with solutions able to minimize the impact of different security attacks.

{\bf Significant Receivers Updates.} Consequently to the previous point, the proposed security techniques have a large impact on the receivers, requiring either hardware or software modification that impacts their operations. When uninterrupted operations should be guaranteed, deploying a new solution for either confidentiality, anti-spoofing, or anti-jamming working at the PHY-layer might find challenges, even if potentially guaranteeing high efficiency at low energy and processing costs.

\textbf{Channel State Information (CSI) Availability.} Looking at information-theoretic schemes for SATCOM confidentiality, being aware of the \ac{CSI} experienced by the passive eavesdropper(s) is critical when calibrating the effectiveness of a security solution. Indeed, if the CSI experienced by the eavesdropper when communicating with the network is available to the transmitter: (i) the secrecy capacity can be maximized; and, (ii) the SNR of the adversary is minimized. Data secrecy and secrecy capacity are handled as constraints of the overall optimization problem, where the overall aim is to ensure that the secrecy rate of the main communication link does not degrade below a minimum threshold or, equivalently, the secrecy outage probability does not exceed a specified upper bound.

{\bf Detection vs. Prevention.} Approaches working on anti-jamming and anti-spoofing mainly focus on detecting the attack once it has been launched. This is a crucial difference from approaches providing confidentiality using information-theoretic schemes that prevent the attacker from gaining information. Such difference is due to the different adversarial models (active in the first two mentioned cases, passive in the second one), that lead to different countermeasures. Combining approaches for multiple security objectives, e.g., anti-jamming and confidentiality, might require new feasibility studies and solutions.

\subsection{Future Directions}
\label{sec:pls_future}
We can identify a few promising future research directions in the area of physical layer security for \ac{SATCOM} as a result of the investigation carried out in this section.

\textbf{Directional Adversarial Antenna.} Almost all the contributions in the three analyzed sub-areas assumed adversaries equipped with omnidirectional antennas. Omnidirectional antennas radiate equal radio power in all directions; thus, they can often be assumed as the worst-case for the integrity of the satellite communications. However, there are specific situations where an adversary equipped with directional antennas can be more disruptive, e.g., in cases where the location of the target communication link is well-known. Directional and semi-directional antennas focus the radiated power in narrow beams, particularly in one direction only~\cite{hurley2007}. From the security perspective, this is an additional powerful feature for an adversary, as it can help to reduce the interference caused by other radio activities, improving the adversary's expected performance. This is an interesting scenario to be investigated for satellite links' security, which has still not been fully explored by Industry and Academia.

\textbf{Security of the Satellite to Satellite Communication Links.} Despite being effective, none of the above-described security solutions provided a security evaluation of the satellite to satellite communication links. This is because of the hardness of both obtaining information about the communication protocols used by such links (often protected by intellectual property rights) and by the nature of such links, envisioned as a kind of core network, far from users' services. 
However, due to the wireless nature of such communications, attacks on these links are both possible and potentially dreadful, as they can disrupt the availability of a \ac{SATCOM} by just affecting the operation of a single link. For instance, Viasat affirms that because LEO satellites are not in constant communication with the ground, a satellite to satellite link can ease the data-sharing mechanism between adjacent satellites~\cite{viasat}. A malicious user could interrupt these types of communications by just jamming such a link. Also, it is unclear if and how physical-layer security techniques can be applied effectively for these links. Thus, investigating the security of Satellite-to-Satellite links is an appealing future research direction.

\textbf{\ac{IRS}.} \acl{IRS} are  one of the most attracting topic in the physical-layer security research community, and they are gaining momentum
~\cite{gong2020_comst}. Thanks to the deployment of a massive number of antenna elements on the satellites, it is possible, in principle, to assist the non-terrestrial network communications by focusing the electromagnetic energy in the intended direction, with consequent benefits in terms of security---receivers out of the intended direction are implicitly excluded from receiving signals~\cite{almohamad2020_ojcs}. A few recent works investigated the theoretical performance of secure communications associated with intelligent reflecting surfaces and the deployment of intelligent reflecting surfaces on UAVs and satellites, e.g., \cite{dong2021_icc}, \cite{xu2021_tvt} and \cite{tekbiyik2020_arxiv}, just to name a few. Therefore, in line with the current trend, we expect a surge of scientific contributions in the upcoming period focusing on the application of IRS for SATCOM scenarios. On the one hand, such works will explore the validity of previous results on PHY-layer security for terrestrial links when deployed to satellite links. On the other hand, when \acp{IRS} are deployed, new security threats might arise, specific to the new technology. Just to provide an example in this direction, an adversary could use a drone flying at significant altitude to boost its SNR when receiving a signal from an IRS deployed on a satellite, or to move within an area with a better coverage without being noticed by terrestrial receivers.

\textbf{GNSS Spoofing Detection via Artificial Intelligence.} In line with a worldwide scientific trend, many recent proposals applied \ac{AI} algorithms to solving GNSS issues, including the detection of spoofing attacks~\cite{simeuri2021_icl}. Specifically, AI-based solutions can be used to detect anomalies in the received signals, so as to identify the presence of the attacker. For instance, the recent proposal by the authors in~\cite{calvo2020_wowmom} use cross-correlation of the received signals to detect anomalous messages, indicating the presence of an attacker. Still, there are several challenges that are to be solved, including the discrimination of legitimate and malicious interference, mobility of the attacker, and use of publicly-available data of satellites (e.g., ephemeris) to imitate satellites movement. In this context, we also notice that several real-world dataset are available and released as open-source, such as \cite{semanjski2019_icl} and \cite{semanjski2020_sensors}.
Thus, we the are seems ripe to experience a renewed interest, likely ignited by 
researchers and industry with expertise in the application of AI algorithms to different domains.

\textbf{Friendly Jamming.} Friendly jamming techniques disrupt all the communications in a given area by allowing, at the same time, legitimate parties to communicate~\cite{tedeschi2020_cst}. Friendly jamming can be achieved in several different ways, e.g., through pre-shared knowledge of jamming time and frequency patterns, or via specific positioning of the legitimate communicating devices, in a way that the \ac{SNR} exceeds the minimum required values only in specific locations, where the legitimate receiving nodes are deployed.

Overall, deploying friendly jamming in SATCOM scenarios may be relevant and useful, e.g., to minimize eavesdropping capabilities of the adversary in military scenarios, while still allowing legitimate devices to communicate in a cheap way.
We notice that none of the security techniques discussed in the previous subsections investigated the feasibility of friendly jamming in the context of \ac{SATCOM}. Note that achieving friendly jamming is not as easy as jamming a communication link. Indeed, friendly jamming requires controlling with extreme precision the timing, the frequency, and the context to be jammed, in a way to know precisely when and how to transmit. In this context, analytical and experimental results about the feasibility and the degrees of freedom of friendly jamming in \ac{SATCOM} are still needed.

\textbf{Spoofing of non-\ac{GNSS} constellations, e.g., NOAA.} At the time of this writing, the majority of contributions dealing with spoofing and anti-spoofing techniques in the \ac{SATCOM} context focused on \ac{GNSS} constellations, due to their higher impact and involvement in everyday life. However, non-\ac{GNSS} constellations such as \ac{NOAA} and \ac{VSAT} are also widely used, e.g., by ships or other devices in remote locations. At the same time, their (low) security is comparable to the one offered by \ac{GNSS} satellites. Solutions to detect and overcome spoofing of the signals emitted by such satellites are still to be investigated and might represent an appealing research opportunity.

\textbf{Powerful Inter-Communications Adversaries.} Most of the approaches using signals from additional communication infrastructures to detect and overcome \ac{GNSS} spoofing and jamming assumed that the adversary only focused on a single communication link. However, a powerful adversary able to inject noise or spoofing signal of \ac{SATCOM} and non-\ac{SATCOM} technologies, eventually using a high powerful antenna, could be disruptive and nullify the effectiveness of such countermeasures. 
For instance, the adversary could fine-tune its attack strategy, so as to deteriorate the performance of such detection strategies. Such powerful adversaries are not too unrealistic to be thought, and thus, are worth investigating.

\section{Cryptography Techniques for SATCOM}
\label{sec:cryptography}

Many contributions have proposed to apply  cryptography techniques to secure SATCOM links. Such works mainly focus on the authenticity and confidentiality of SS and SG communications, and adapt security primitives originated from other domains to work efficiently with \ac{SATCOM} systems. We can notice that part of the scientific contributions in this area adapt the implementation and network architecture of well-known cryptographic solutions to the SATCOM scenario, while the other part study the effectiveness and consequences of the introduction of novel paradigms, such as quantum computing, with an eye on the requirements of SS and SG communication links. 
In this section, we review and classify contributions dealing with the application of cryptography schemes in \ac{SATCOM}, classifying them based on the provided security service. Section~\ref{sec:authentication} focuses on techniques for peer authentication, Section~\ref{sec:ka} discusses key agreement schemes, while Section~\ref{sec:kd} introduces approaches for key distribution based on quantum channels. The main lessons learnt from our study are reported in Section~\ref{sec:crypto_lessons}, while Section~\ref{sec:crypto_future} outlines promising research directions in this domain.

\subsection{Authentication}
\label{sec:authentication}

SATCOM systems involving users, mobile devices, and ground stations (or network control centers) require establishing trust among the cited entities. However, due to the presence of the wireless medium, SATCOMs are more prone to impersonation attacks. To mitigate this problem, various authentication protocols have been proposed in the literature. Table~\ref{tab:authentication} provides a comprehensive classification of the most important scientific contributions in the field, cross-comparing them across the selected communication architecture, the proposed cryptographic technique, the security properties, the security analysis, and the assessment methodology.

\begin{table*}[htbp]
	\centering
	\scriptsize
	\caption{Comparison of Different Authentication Methods for SATCOM links.}
		\begin{tabular}{|P{0.5cm}|P{3.1cm}|P{2.5cm}|P{3.2cm}|P{1.2cm}|P{1.5cm}|}
		\hline
        \textbf{Ref.} & \textbf{Link} & \textbf{Key Sharing Technique} & \textbf{Security Properties} & \textbf{Security Analysis} & \textbf{Assessment} \\
        \hline
        \cite{ibrahim_2016_scn} & User to Network Control Center through Satellite & ECC & Mutual Authentication, User Anonymity, Unlinkability, Non-Repudiation & Formal & Simulations \\
        \hline
        \cite{chen_2014_amis} & User to Network Control Center through Satellite & Pre-Shared Key & Mutual Authentication, User Privacy, Minimum Trust & Informal & \xmark \\
        \hline
        \cite{xinghua_2017_isc} & User to Network Control Center through Satellite & Pre-Shared Key & Mutual Authentication, User Privacy, Minimum Trust & Formal & Simulations \\
        \hline
        \cite{xu_2020_ijscn} & User to Network Control Center through Satellite & ECC & Mutual Authentication, User Anonymity, Untraceability & Informal & Informal \\
        \hline
        \cite{zhang_2015_ijscn} & User to Network Control Center through Satellite & Pre-Shared Key & Mutual Authentication, User Privacy, Minimum Trust & Informal & \xmark \\
        \hline
        \cite{lin_2016_ijscn} & Satellite to Network Control Center & Pre-Shared Key & Mutual Authentication & Informal & \xmark\\
        \hline
        \cite{zhao_2016_milcom} & User to Network Control Center through Satellite & Pre-Shared Key & Mutual Authentication & Formal & \xmark\\
        \hline
        \cite{liu_2017_ijscn} & User to Network Control Center through Satellite & Pre-Shared Key & Mutual Authentication, User Privacy, Minimum Trust & Discussion, Formal & \xmark \\
        \hline
        \cite{caparra_2016_iclgnss, caparra2016novel} & \ac{GPS}/\ac{GNSS} Satellites & TESLA & Message Authentication & Informal & Simulations \\
        \hline
        \cite{hernandez_2014_ion-gnss} & Galielo/\ac{GNSS} Satellites & TESLA & Message Authentication & Informal & Simulations \\
        \hline
        \cite{kerns2014blueprint} & \ac{GPS}/\ac{GNSS} Satellites & PKC and TESLA & Message Authentication & Informal & Simulations \\
        \hline
        \cite{huang2020mutual} & SS, SG Control Center & Symmetric Encryption & Mutual Authentication, Message Authentication & Informal & Experiments \\
        \hline
        \cite{jurcut2019novel} & User to Network Control Center through Satellite & Pre-Shared Key & Mutual Authentication, User Privacy& Informal & \xmark \\
        \hline
        \cite{ghorbani2020navigation} & \ac{GPS}/\ac{GNSS} Satellites & TESLA & Message Authentication & Informal & Simulations\\
        \hline
        \cite{curran_2014_encgnss} & Galielo/\ac{GNSS} Satellites & TESLA & Message Authentication & Informal & Experiments \\
        \hline
        \cite{meng2018low} & User to Ground Station through Satellite & ECC & Unforgeability, Mutual Authentication, Conditional Anonymity & Formal & Experiments\\
        \hline
	    \end{tabular}
	\label{tab:authentication}%
\end{table*}

\textbf{Link.} Most of the analyzed works considered the User to Ground Station/Network Control Center through Satellite link depicted in Figure~\ref{fig:architecture}, applying standard cryptography techniques to secure the channel. Many schemes, i.e., \cite{caparra_2016_iclgnss,caparra2016novel,hernandez_2014_ion-gnss,kerns2014blueprint,ghorbani2020navigation} and \cite{curran_2014_encgnss}, considered the protection of the \ac{GPS}/\ac{GNSS} communication link, while other links are rarely taken into account.

\textbf{Key Sharing Technique.} Many different key sharing techniques are used for equipping the entities with the crypto material necessary to run authentication protocols. 
The authors in~\cite{ibrahim_2016_scn, xu_2020_ijscn, meng2018low} proposed a public key cryptographic scheme based on \ac{ECC} to provide peer entity authentication. Conversely, other proposals assume a pre-shared key, statically known by one or more users.
Some other approaches, i.e., \cite{caparra_2016_iclgnss, caparra2016novel, hernandez_2014_ion-gnss, kerns2014blueprint, ghorbani2020navigation, ghorbani2020navigation, curran_2014_encgnss}, adopt the Timed Efficient Stream Loss-tolerant Authentication (TESLA) protocol conceived by Perrig \textit{et. al} in~\cite{perrig2002tesla} to provide delayed source authentication for broadcast communications in very resource-limited environments, by leveraging only symmetric cryptographic primitives and hash chains.
\textcolor{black}{Focusing on schemes adopting asymmetric cryptography solutions, the adoption of \ac{ECC} rather than the traditional RSA provides benefits in terms of smaller public and private keys for the same security level, faster key generation and signature operations, as well as low overhead on CPU and memory usage. \ac{RSA} schemes are usually very simple to implement, widely deployed in the industry, and specific public key operations such as the signature verification are usually faster than the ones on \ac{ECC}, considering the same size of the elements. 
However, the setup of a public-key infrastructure to manage, distribute, and revoke public key certificates is costly and time-consuming. Thus, when a public key infrastructure is not desirable or affordable, and the efficiency of the crypto operations is at premium, symmetric cryptography solutions for digests generation are adopted, such as the TESLA protocol \cite{perrig2002tesla}. Symmetric solutions  allow to execute very fast encryption and decryption operations, as well as to generate authentication digests that can be produced and verified very quickly. However, differently from public-key solutions, solutions based on symmetric cryptography requires the communicating parties to share a secret, to be kept private at least for a given amount of time (e.g., in case of TESLA). 
If a shared  key 
is compromised, it should be discarded and replaced. However, replacing and updating a key 
can be a time-consuming activity, especially in a context where the communicating entities are orbiting several kilometers above the Earth surface.
}

\textbf{Security Properties.} The main security property provided by the proposed schemes is mutual authentication, i.e., entities authenticate each other before establishing mutual communication. Equal importance is given to message authentication, which ensures that the message was actually sent by the entity that claimed to have done so. Contributions such as~\cite{ibrahim_2016_scn,xu_2020_ijscn,meng2018low,chen_2014_amis,xinghua_2017_isc,zhang_2015_ijscn,liu_2017_ijscn,jurcut2019novel} guarantee additional properties, such as anonymity and user privacy, adopting techniques that hide the user identity during a communication.

\textbf{Security Analysis.} The schemes proposed by the authors  in~\cite{ibrahim_2016_scn,zhao_2016_milcom,hernandez_2014_ion-gnss,meng2018low} are formally proved as secure with reference to certain formal specifications or properties. 
Some tools that help to prove these properties are ProVerif~\cite{proverif}, CryptoVerif~\cite{blanchet2017cryptoverif}, AVISPA~\cite{avispa}, and Tamarin~\cite{tamarin}, to name a few. Note that only the security of the cryptographic scheme can be verified through this mechanism, while its integration in the reference system architecture could widen the threat surface.

\textbf{Assessment Methodology.} Similarly to the works dealing with physical-layer security, most of the analyzed schemes use simulation-based evaluation. Only few of them, i.e., \cite{huang2020mutual,ghorbani2020navigation,curran_2014_encgnss,meng2018low}, used real data and deployed proof-of-concept.

\subsection{Key Agreement}
\label{sec:ka}

Key agreement protocols (a.k.a. key establishment protocols) are used to allow two (or possibly more) entities that could not have anything in common to agree on a shared key to be used to secure further mutual communications~\cite{stallings2012_book}. Nowadays, key agreement protocols via public key cryptography or pre-shared keys are used in a range of different security protocols. Although several works proposed lightweight key establishment solutions integrating well-known cryptographic approaches in a variety of application domains, key establishment mechanisms in \ac{SATCOM} have received only reduced attention. Table~\ref{tab:keyagreement} provides a comprehensive classification of the scientific contributions dealing with key agreement in \ac{SATCOM} scenarios, considering reference system requirements and features.
\begin{table*}[htbp]
	\centering
	\scriptsize
	\caption{Comparison of Different Key Agreement Methods for SATCOM links.}
		\begin{tabular}{|c|P{1.5cm}|P{1.8cm}|P{2.5cm}|P{2.1cm}|P{2.5cm}|}
		\hline
        \textbf{Ref.} & \textbf{Link} & \textbf{Cryptography Technique} & \textbf{Target Security Service} & \textbf{Adversary Model} & \textbf{Assessment} \\
        \hline
        \cite{ostadsharif_2019_comcom} & SG & ECC & Authentication, Confidentiality, Integrity & Canetti–Krawczyk & Simulations and Experiments on Smartphone \\
        \hline
        \cite{murtaza2019lightweight} & General Purpose, SG, SS & ECC, RSA & Authentication, Integrity & Active & Formal Analysis \\
        \hline
        \cite{caparra2017key} & SG (\ac{GNSS}) & ECC, RSA & Authentication, Integrity & Active & Simulations \\
        \hline
        \cite{deng_2018_iaeac} & SG (Beidou) & RSA & Authentication, Integrity & \xmark & Experimental (Ground Part) \\
        \hline
        \cite{altaf2020lightweight} & SG, SS & Pre-Shared Key & Authentication, Confidentiality, Integrity, Anonymity & Active/Passive & Formal Analysis, Simulations\\
        \hline
        \cite{yantao_2010_jcn} & SG, SS & Identity Based Cryptography & Authentication, Confidentiality, Integrity & extended Canetti–Krawczyk & Formal Analysis \\
        \hline
        \cite{lee_2013_ijscn} & SG (VSAT) & Chaotic Maps & Authentication, Confidentiality, Integrity, Availability  & Active & Complexity Analysis \\
        \hline
        \cite{qi_2019_ijscn} & SG & Pre-Shared Key, ECC, RSA & Authentication, Confidentiality, Integrity, Availability & Active & Formal and Complexity Analysis\\
        \hline
	    \end{tabular}
	\label{tab:keyagreement}%
\end{table*}

\textbf{Link.} Most of the presented approaches focus on the key establishment in SG and SS links. It is also worth noting that most of the solutions require a software update that can be done via radio link or, in particular situations, by intervening offline on the satellite.

\textcolor{black}{
\textbf{Cryptography Technique.} Many different techniques are used to allow the entities to derive a shared secret. Overall, the same considerations introduced for the key sharing techniques in Section~\ref{sec:authentication} are still valid for key agreement protocols adopted in SATCOM links. For completeness, we introduce the pro and cons also for the Identity Based Cryptography and the Chaotic Maps. In Identity Based Encryption (IBE) schemes, each communicating entity owns a unique identifier, adopted to compute the correspondent public key. With such an approach, no certificates are needed and no pre-enrollment is required. However, the architecture requires a key generation center, which might be vulnerable to key-escrow attacks, and therefore, exposed to a risk of information disclosure~\cite{joye2009_ide}. Unlike public-key cryptography systems, chaotic maps do not require modular arithmetic, being therefore very fast 
for both encryption and digital signature. Moreover, algorithms based on chaotic maps might not require that large private keys, being computationally efficient. On the downside of such a technique,  chaotic maps can produce ciphertext that are bigger than the plain-text~\cite{kocarev2001_csm}.
}

\textbf{Target Security Service.} Most of the presented schemes focus on the balanced protection of the Confidentiality, Integrity, and Availability of data (CIA) by providing, at the same time, identity verification. 

\textbf{Adversary Model.} The Canetti–Krawczyk (CK) and the extended Ca\-netti–Krawczyk (eCK) security models~\cite{sarr2010_springer}, are widely used to verify and provide the aforementioned security properties for key agreement protocols. These models have been developed to build secure protocols that guarantee peer entity authentication and message authenticity during the key exchange procedure. Indeed, the main aim of these models is to provide a method so that the security protocols proposed in the literature can have a match between the implementation and the adoption in a real environment by taking into consideration also possible attacks of an active adversary~\cite{bellare1998modular}.


\textbf{Assessment Methodology.} In order to evaluate the performance of a key agreement protocol, authors can evaluate the offered security properties by using the formal analysis or discuss the solution. Further, they can estimate the efficiency via simulation tools, like~\cite{ostadsharif_2019_comcom,caparra2017key} and \cite{altaf2020lightweight}, or by performing a complexity analysis such as~\cite{lee_2013_ijscn,qi_2019_ijscn}. The only contributions providing performance on a real system, i.e., \cite{ostadsharif_2019_comcom,deng_2018_iaeac}, worked on the ground link due to the viability of the approach and its 
reduced costs.

\subsection{Quantum Key Distribution}
\label{sec:kd}
\begin{figure}[H]
  \centering
  \includegraphics[angle=0, width=\columnwidth]{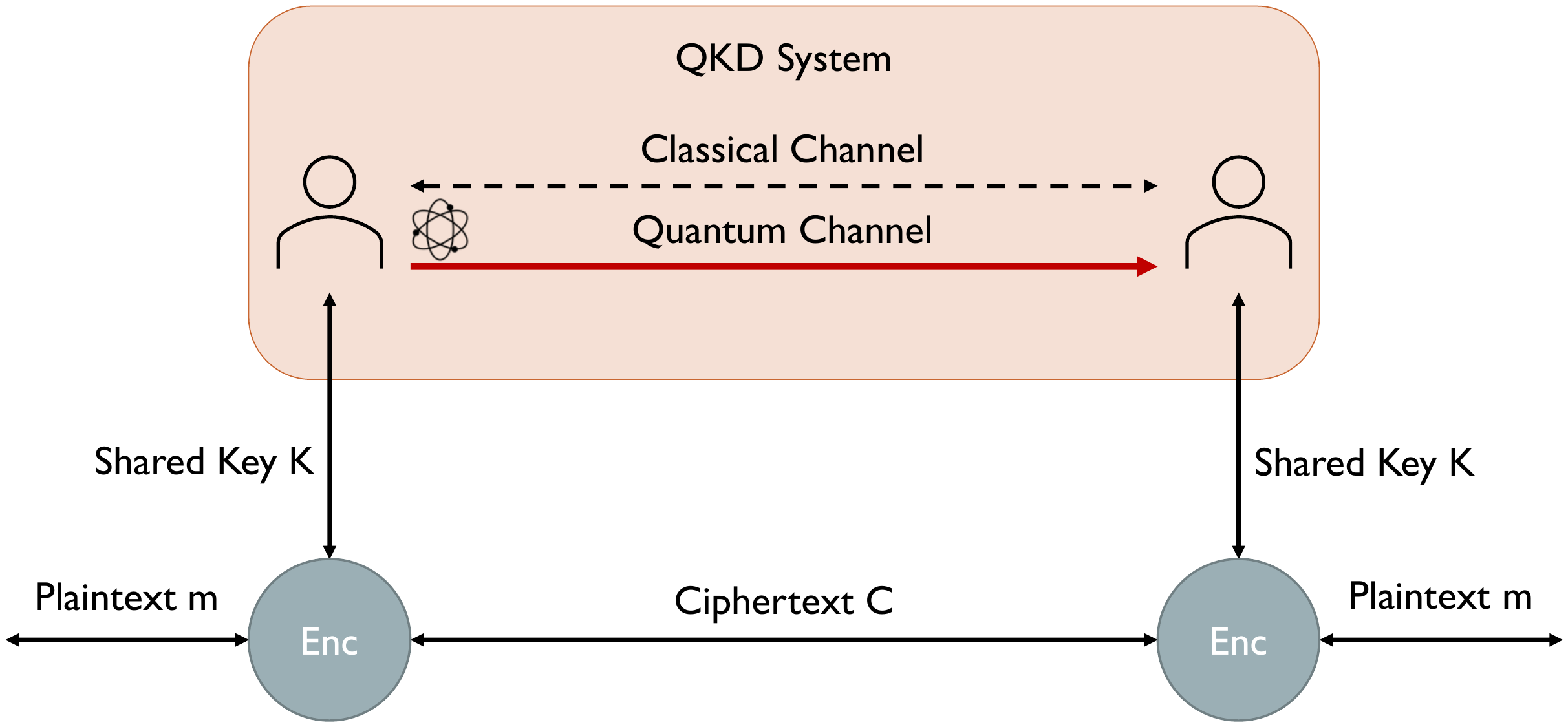}
  \caption{Generalization of the QKD system architecture.}
  \label{fig:qkd_architecture}
\end{figure}

Although the soundness of the adopted encryption techniques typically relies on traditional mathematics proofs, quantum architectures are coming out of laboratories to be used in many contexts, based on different assumptions.
Overall, Quantum key distribution (QKD) allows two remote parties to securely negotiate a cryptographic key even in the presence of an eavesdropper. However, compared to traditional key distribution schemes, the security of such distribution mechanisms does not rely on cryptographic assumptions (i.e., difficulty of solving specific mathematical problems), but on the unique quantum mechanics properties of the adopted communication strategy at the physical-layer. Thus, even assuming a powerful adversary with unlimited computational capabilities and able to break cryptography assumptions, the robustness of the protocols still holds---
basically, QKD protocols allow detecting the presence of an eavesdropper on the communication link~\cite{diamanti2016_qi}.\\
Currently, several \ac{QKD} solutions have been proposed in the literature, even if there is still a gap between the information theory and practical implementations.
As depicted in Fig.~\ref{fig:qkd_architecture}, \ac{QKD} systems work by using photons, i.e., particles which transmit light to transfer data~\cite{gyongyosi2019_csr}. 
Quantum technology allows two distant entities to agree on a common symmetric key even if they do not share any previous knowledge. The key is adopted with the respective encryption algorithm to transmit and receive encrypted messages over a standard communication channel, even if the \ac{OTP} encryption scheme is the most used one.
The benefit of this ``unbreakable'' encryption is that the data is carried via photons, which cannot be copied or eavesdropped without leaving evidence of such an attempt. Indeed, an adversary measuring the state of a 
photon would disturb the channel, hence  compromising the key agreement procedure and providing a kind of tamper-detection evidence. In terms of security, quantum computing 
could make the current state of the art on security obsolete, jeopardizing the protection of data and communications. This aspect is leading to an acceleration of the adoption of countermeasures (e.g. post-quantum encryption algorithms), especially to protect data and critical infrastructures such as SATCOMs~\cite{khan_2018_osa, bedington_2017_nqi}. Table~\ref{tab:keydistribution} provides an overview of the most important \ac{QKD} techniques proposed in the context of \ac{SATCOM}, and cross-compares them across reference system features and requirements.

\begin{table*}[htbp]
	\centering
	\scriptsize
	\caption{Comparison of Different \ac{QKD} Methods proposed for \ac{SATCOM} applications.}
		\begin{tabular}{|P{0.5cm}|P{2.5cm}|P{3.5cm}|P{4cm}|P{1.5cm}|}
		\hline
        \textbf{Ref.} & \textbf{Link} & \textbf{Protocol} & \textbf{Main Contribution} & \textbf{Assessment} \\\hline
        \cite{benton_2010_oc} & SG, SS & BB84 & Free Space QKD & Experiments\\\hline
        \cite{tomaello_2011_asr} & SG (GEO, LEO), SS & BB84/BB84 + Decoy/B92 & Estimate Link Attenuation and SNR & Simulation\\\hline
        \cite{mafu_2013_pra} & Generic & Entanglement-Based & Efficiency and Security Analysis & Experiments\\\hline
        \cite{vallone2015experimental} & SG (LEO) & BB84 & Feasibility of QKD  & Experiments \\\hline
        \cite{tan_2015_jmo} & SG (LEO) & Generic QKD & Radiation Tolerance estimation of single photon detector & Simulation\\\hline
        \cite{bourgoin_2015_pra} & SG (LEO) & BB84 + Decoy & Efficiency of QKD in high-loss regime scenarios & Experiments\\\hline
        \cite{liao_2017_nature} & SG (LEO) & BB84 + Decoy & QKD with Quantum Repeaters & Experiments\\\hline
        \cite{takenaka_2017_nature} & SG & B92 & Communication Feasibility of QKD & Experiments\\\hline
        \cite{toyoshima_2011_ijo} & GS & B92 & Prototype of free-space QKD & Experiments\\\hline
        \cite{jennewein_2014_etsd} & SG & Custom QKD & Long Distance QKD & Experiments\\\hline
        \cite{sharma_2018_icccnt} & SG (LEO) & Generic QKD & Quantum Error Correction Technique & Simulation\\\hline
        \cite{bonato_2009_njp} & SG (LEO) & BB84 $\pm$ Decoy/Ekert91 & Efficiency \& Performance Analysis & Simulation\\\hline
        \cite{bedington_2015_small} & SG & Entanglement-based QKD & Miniaturised entangled photon sources & Experiential\\\hline
        \cite{wang_2013_nature} & SG (LEO) & Decoy-State QKD & Performances Analysis on High-loss & Experiments\\\hline
        \cite{liao_2017_nature2} & SG (LEO), SS (LEO) & Custom QKD & Feasibility of satellite-based quantum communication in daylight & Experiments\\\hline
        \cite{rarity_2002_njp} & GS (LEO) & BB84 & Securing Key Exchange Sat to Ground & Experiments\\\hline
	    \end{tabular}
	\label{tab:keydistribution}%
\end{table*}

\textbf{Link.} Most of the analyzed contributions propose approaches leveraging a quantum technology for SG and SS communication links. Note that such communication links might be challenging to manage when considering interference, natural phenomena such as sunlight, and the long distances among the entities.

\textcolor{black}{
\textbf{Quantum Protocols.} \ac{QKD} protocols are today already adopted to protect communications through optical communication channels. QKD systems are already operational in different contexts, also allowing long-distance connections in \ac{P2P} communications. Several works in the literature such as \cite{benton_2010_oc,tomaello_2011_asr,vallone2015experimental} and \cite{bonato_2009_njp} adopt the well known BB84 quantum key distribution scheme, developed by Charles Bennett and Gilles Brassard in 1984. BB84 was originally described using photon polarization states where no quantum entanglement was required. Conversely, the authors in \cite{mafu_2013_pra}, and \cite{bedington_2015_small} propose schemes based on quantum entanglement, i.e., a particular phenomenon where particles remain intimately connected, even if separated by long distances~\cite{chen2021nature}. When it comes to \ac{QKD}, while the high-level protocols are well understood and are taught in university courses, and commercial products are available on the market~\cite{quantropi}, the underlying phenomena that make them possible are rooted on the evolving frontiers of physics~\cite{takeda2021quantum}. As such, a detailed discussion of the cited protocols is beyond the scope set for this survey. However, interested readers can resort to recent authoritative surveys on \ac{QKD} available in the literature, such as~\cite{miralem2021_csur, sharma2021_ojcs, xu2020_rmp}, to cite a few.
}

\textbf{Main Contribution.} A large variety of contributions used to distribute the cryptographic key via \ac{QKD}. Approaches use either free space \ac{QKD} (\cite{benton_2010_oc,toyoshima_2011_ijo}), or perform feasibility and efficiency analysis with reference to some specific schemes (\cite{vallone2015experimental,mafu_2013_pra,bourgoin_2015_pra,bonato_2009_njp,takenaka_2017_nature,liao_2017_nature2}). All these techniques share the basic consideration that the quantum technology cannot be compromised due its intrinsic properties.

\textbf{Assessment Methodology.} Differently from approaches based on classical cryptography, most of the contributions focusing on QKD carry out also an experimental assessment, allowing the authors to demonstrate the feasibility of satellite-based quantum communications by experimentally analyzing their efficiency, error tolerance, and security properties.

\subsection{Lessons Learned}
\label{sec:crypto_lessons}

In the following, we summarize the most important lessons learned from the analysis of cryptography approaches for SATCOM reported in the previous subsections.

{\bf Software Modifications.} Differently from approaches based on PHY-layer security, cryptography solutions always require the modification of the software running on the receivers and on the satellites. Some approaches actually provided the required modifications, while others assume that the advantages of such modifications overcome the cost and effort required to install them.

{\bf Impact of Software Updates in \ac{SATCOM}.} Cryptography-based solutions always require a dedicated software update on the satellite, and this could also affect the operational status of the \ac{SATCOM} communication link. Thus, the cost of their integration should be carefully taken into account, and they should be applied only when other solutions (e.g., PLS-based ones) do not guarantee sufficient security.

{\bf Quantum Key Agreement for \ac{SATCOM}.} Due to its promising level of security, quantum computing strategies show appealing advantages for \ac{SATCOM} links, and researchers have already started to evaluate its feasibility due to the large involved distances~\cite{chen2021nature}. We expect to see many contributions to come on this topic in the following years.


\subsection{Future Directions}
\label{sec:crypto_future}

\textbf{Analysis of Security Requirements.} Most of the works considered in the first two subsections of this section only apply well-known cryptography schemes in \ac{SATCOM}, plugging them in without a strong motivation or a detailed description of the underlying security requirements. 
Due to the significant impact that cryptography has on the operation and performance of \ac{SATCOM} deployments, researchers and industry should come up with a dedicated security analysis of \ac{SATCOM} links, explaining precisely what the threats are and why cryptography solutions are advantageous compared to Physical-layer ones in solving such issues. To the best of our knowledge, such a study is still not available in the literature.


\textbf{Communication Channel Availability.} In the presence of an eavesdropper, a quantum-based communication channel is disrupted, and the parties cannot continue to communicate. In principle, this capability could help to detect \ac{MITM} attacks and identify potential eavesdroppers quickly. However, it also paves the way for easy \ac{DoS} attacks, even harder to detect in the context of \ac{SATCOM} due to the large reception range and coverage area of the communication. In this context, backup solutions for guaranteeing the availability of the service are needed, as well as tools to discriminate if the channel if compromised because of an eavesdropper, or because of the channel noise.

\textbf{Quantum Channel Security Assessment Tools.} At the time of this writing, no works are available that evaluate the physical-layer security of \ac{QKD} signal generation tools. In principle, passive side-channel attacks can be conducted to extract meaningful information from a \ac{QKD} channel, without affecting and compromising the robustness provided by \ac{QKD} strategies. It is essential to explore this research area to provide methods and tools to mitigate this open issue~\cite{li2021physrev}.

\section{Emerging Research Challenges}
\label{sec:challenge}

The previous sections delved into the most active research branches related to the SATCOM domain and provided some appealing future research directions in those specific contexts. However, in addition to the identified research areas, our investigation highlighted further security-related \ac{SATCOM}-based application domains that are receiving increasing attention from the scientific and business community. In the following, we discuss some of them, showing their key challenges and potential to attract additional interest in the years to come.

\textbf{Cognitive Satellite Terrestrial Networks.} Cognitive radio in the context of wireless communication systems is a research area that attracted lots of interest in the few last years. In a nutshell, cognitive radio systems allow the coexistence of primary users (using devices that own the license to use a specific frequency band) and secondary users (allowed to share resources with the primary network, but not in possession of the license) on the same network and spectrum, sharing the same radio resources. A few works applied the concept of cognitive radio networks in the context of \ac{SATCOM}. For instance, the authors in \cite{an2016_jsac, lin_2018_jsac, li_2018_ieeewcl, li_2018_tvt} propose to secure the communication in cognitive satellite-terrestrial networks. They assume a scenario where the primary network is constituted by GEO, MEO, or LEO satellites, sending confidential messages to the fixed-satellite operator in the presence of eavesdroppers (secondary users) attempting to capture the satellite information signal. In their own (secondary) network, the network operator communicates with the user terminals. Still, doubts are there on the actual applicability of cognitive radio techniques in the context of satellite-terrestrial networks. This is mainly because of the extremely wide coverage of satellites, where the CR techniques might not work well.  In line with a large amount of work done in the context of cognitive radio for terrestrial networks~~\cite{salahdine2020security}, we expect increased interest in this domain in the next years.

\textbf{Drone-To-Satellite.} 
\ac{UAV}, a.k.a. drones, have gained increased momentum in the last years, in both academia and Industry~\cite{mozaffari2019_comst}. In the context of \ac{SATCOM}, one of the most critical challenges consists of allowing secure communication between small/commercial UAVs and satellites.
\textcolor{black}{For instance, the authors in~\cite{yin2022_tits} proposed a physical layer security framework in space-air-ground (SAGIN) downlink multi-beam satellite-enabled vehicle communications, where the UAV is adopted as cooperative node, interacting with the legitimate user and acting as a source of artificial noise to mitigate eavesdropping. In the same network setup, the authors in~\cite{cheng2019_jsac} investigated the IoT computing offloading problem by proposing a reinforcement learning approach  to allocate the resources of the UAV edge server efficiently. In the same domain, the authors in~\cite{zhang2017_commag} proposed a software defined architecture supporting different vehicles in an efficient manner. 
In line with existing works such as~\cite{emilien2021}, we forecast numerous appealing applications involving drones and satellites. Using satellites links, users can: (i) drive drones remotely; (ii) stream video from the drone's camera; (iii) use the drone to collect information from remote satellites; and, (iv) use the drone for optical remote sensing applications. Due to the well-known security, safety, and privacy issues posed by drones usage, and due to the central role of drones in the development of the upcoming 6G communication systems~~\cite{ray2021_ksu}, we expect significant research activity in this domain in the years to come.}

\textbf{\ac{AI} in \ac{SATCOM}.} The usage of \ac{AI}-based techniques is gaining increasing importance  in almost any application domain, cybersecurity included. In the context of \ac{SATCOM}, \ac{AI} techniques could be used for many purposes, e.g., to identify physical-layer characteristics of the signals emitted by the satellites, to discriminate between authentic and injected signals, and for intrusion detection, to name a few . In this context, the authors in~\cite{oligeri2020past} experimentally show that using a dedicated Convolutional Neural Network (CNN) it is possible to fingerprint the raw IQ samples received from LEO Satellites (Iridium) and authenticate the emitting transceiver on board of the satellite, despite the large distances. We expect increasing attention toward this research domain, targeting additional satellites constellations (\ac{GNSS} ones included) or other applications of AI.

\textbf{Software-defined satellites.} The integration of \ac{SDN} into \ac{SATCOM} could improve the connectivity coverage and performance for using broadband communications by allowing the operators to reconfigure the satellites as needed~\cite{rechenberg2021_icmcis,papa2020_tnsm,bertaux2015_commag}. However, \ac{SDN} also come with their security issues, that are further specialized in \ac{SATCOM} use-cases~\cite{eliyan2021_fgcs}. Additional research is needed in this context.

\textbf{Network Slicing for the Internet of Space Things.} The continuous development of nano-satellites is accelerating the deployment of low-cost satellite networks~\cite{ahan2021_tnsm}. Emerging paradigms, such as the \ac{IoST}, require a network slicing framework to provide the support for the plethora of space-application scenarios. A network slice is a part of the network that is independent and logically separated from the rest. A specific slice has specific security policies, used to protect the slice while meeting specific system requirements. However, there are neither  common strategies nor protocols suitable to design a network slice in the context of \ac{SATCOM}. Despite initial studies in this context are available~\cite{akyildiz2019_ieeenet,akyildiz2019_comnet}, major work is  still to be done, and we expect increasing attention towards this topic.

\textbf{Green Satellites.} The design of environmentally-friendly satellites can help to reduce the environmental impact of a satellite, its production cost, and maintenance compared to traditional ones. However, reducing the cost and the impact of the satellite inevitably could affect the provided security services. This emerging research area, also suggested from the \ac{ESA}~\cite{greensat}, leads to a potential redesign of the existing procedures and technologies, also including the security domain. Definitively, a novel and  interesting research topic.

\textbf{Satellites Signals for Opportunistic Navigation.} Specific satellites signals can be used to pinpoint a specific location on Earth, similar to the \ac{GPS}. A group of researchers developed a working solution based on the cited logic, leveraging signals broadcasted by Starlink internet service satellites~\cite{kassas2021_taes}. The usage of additional satellite constellations could provide reliability and spoofing detection mechanisms for devices on Earth, and more research into the robustness of such solutions is needed.

\textbf{Cybersecurity for Commercial Satellite Operations.} The \ac{NIST} is seeking comments on the draft specification NISTIR 8270, which describes the security procedures and the concepts for commercial space operations. The draft considers the management aspects, risk management operations, and defines the requirements that ``might coexist within space vehicle systems''. The \ac{NIST} is requiring feedback on the overall approach, the example use case, and the identified controls for the proposed use case~\cite{Scholl2021}. We expect that several research contributions could come out due to the study and application of this (yet to come) recommendation to real use-cases. 

\textcolor{black}{
\textbf{Standardization of Security for Non-Terrestrial Networks.} In the standardization community, and in particular, within the Third Generation Partnership Project (3GPP) committee, satellite communications are specifically considered in the design of the Non-Terrestrial Networks (NTNs), anticipating the upcoming tight integration between terrestrial, aerial and satellite networks~\cite{rinaldi2020_access}. Specifically, the standardization of Non-Terrestrial Networks (NTNs) has been launched by 3GPP in the 3GPP Release 16~\cite{3gpp_rel16}. At the same time, new security aspects have been recently defined by the 3GPP in the Release 17~\cite{3gpp_rel17}, and new amendments are planned in the upcoming Release 18~\cite{3gpp_rel18}. \\
In this context, none specifications edited by the 3GPP specifically took into account network security issues for NTNs. As a result, the current approach recommended by the 3GPP consists of a straightforward integration of the 5G security architecture and protocols into NTNs. Such an integration, however, comes with several challenges, in terms of communication overhead, software updates, and unreliability of the wireless links. Specifically, security issues in the operation of NTNs have been investigated by the Working Group on Satellite 5G, established within the IEEE Future Networks Initiative~\cite{kota2021_icc}. In one of the latest deliverables of the WG, i.e.,~~\cite{deliv_sat5g}, they drew a roadmap of the priorities to be addressed in this regard, highlighting that new security mechanisms might be needed for specific deployments, and that the emphasis should be put on the isolation of the end-users from the shared NTN network. The WG realized several security-related activities, i.e.,: (i) analyzed the state of the art about security for NTNs; (ii) provided a threat analysis for the NTN scenario; (iii) identified specific complications derived from using of 5G security solutions on 5G-NTN networks, by experimentally verifying them on prototyping platforms; and, finally, (iv) identified additional security concerns, mainly related to the integration of emerging technologies such as network slicing, edge computing, and multicasting over satellite networks (see Section 2.6.2 of~\cite{kota2021_icc}. The temporary recommendation proposed by the WP was to adopt the IPSec protocol suite to secure the communication link, but they also recommended further study into the issues at the 3GPP standardization level. However, the 3GPP refused to investigate further into the issue, at the time of this writing, still recommending a straightforward integration of 5G security into the NTN domain~\cite{deliv_sat5g}.  \\
Nonetheless, due to the forecasted performance issues arising from such the integration of 5G-security into NTNs, we expect significant contributions by the research community in the years to come, potentially triggering dedicated and ad-hoc initiatives by the 3GPP.
}

\textbf{Security and Privacy for 6G.} 6G networks will accommodate satellites, UAVs, and undersea communications~\cite{wang2020_dcn}. It is crucial that any security proposal framed in this context protect the communications while guaranteeing reliability, low latency, and secure and efficient transmission services. Physical-layer security is the first candidate defense for these new emerging technologies, but emerging cryptography-based solutions could also play a role if their integration is carefully systematized and orchestrated with the existing services. \textcolor{black}{In the context of 6G initiatives, the 3GPP claimed that for the next few years (2030s) additional research is needed into this application area. Adapting and integrating the security services on satellites with mobile terrestrial/sea systems while meeting the requirements of 6G communication services will indeed represent a complex and difficult challenge~\cite{nguyen2021_comst}. For the cited scenario,  we expect the adoption of real-time security communication protocols and emerging architectural solutions, such as Zero Trust~\cite{rose2020zero}.}


\section{Conclusion}
\label{sec:conclusion}

In this contribution, we have provided a survey of the most significant link-layer security issues, threats, and mitigation techniques adopted in the context of \aclp{SATCOM} systems. First, we presented  general background on the \ac{SATCOM} architecture, the most important constellations, and network parameters. Then, we divided the relevant literature on the topic into two major research areas, i.e., physical-layer security and cryptography, and we further identified dedicated topics in each macro-area, focusing on specific threats. For the physical-layer area, we discussed and cross-compared solutions based on the usage of information-theoretic security schemes, anti-jamming strategies, and anti-spoofing schemes. For the cryptography area, we specifically discussed approaches for authentication, key agreement, and key distribution based on the emerging quantum computing paradigm. We also identified lessons learned and specific future directions for each of the cited threats and research areas. Finally, we presented a few appealing emerging challenges in the \ac{SATCOM} security domain, pointing out the main research challenges to be solved and the areas where new contributions from the scientific community might have major impact. 

Overall, we believe that the exposed research challenges highlight 
that the design and testing of cybersecurity strategies for SATCOMs is still an active research domain. In particular,
our contribution calls for collaboration between Industry and Academia to
unlock new business opportunities and services, while enjoying the needed level of security for communications, applications, and infrastructures.

\section*{Acknowledgements}
The authors would like to thank the anonymous reviewers, that helped improving the quality of the paper.\\
This publication was partially supported by the Technology Innovation Institute, Abu Dhabi - UAE, and awards NPRP-S-11-0109-180242 from the QNRF-Qatar National Research Fund, a member of The Qatar Foundation, and NATO Science for Peace and Security Programme - MYP G5828 project ``SeaSec: DronNets for Maritime Border and Port Security''. This work has been partially supported also by the INTERSCT project, Grant No. NWA.1162.18.301, funded by Netherlands Organisation for Scientific Research (NWO). The findings reported herein are solely responsibility of the authors.

\balance
\bibliographystyle{elsarticle-num}
\bibliography{biblio}

\begin{thebibliography}{100}
\expandafter\ifx\csname url\endcsname\relax
  \def\url#1{\texttt{#1}}\fi
\expandafter\ifx\csname urlprefix\endcsname\relax\def\urlprefix{URL }\fi
\expandafter\ifx\csname href\endcsname\relax
  \def\href#1#2{#2} \def\path#1{#1}\fi

\bibitem{maral2020satellite}
G.~Maral, M.~Bousquet, Z.~Sun, {Satellite communications systems: systems,
  techniques and technology}, John Wiley \& Sons, 2020.

\bibitem{spacex_starlink}
Space.com, \href{Starlink: SpaceX's satellite internet project}{{Starlink:
  SpaceX's satellite internet project}}, (Accessed: 2022-Jul-10) (2021).
\newline\urlprefix\url{Starlink: SpaceX's satellite internet project}

\bibitem{theverge_facebook_amazon}
J.~Porter,
  \href{https://www.theverge.com/2021/7/14/22576788/amazon-acquires-facebook-satellite-team-project-kuiper}{{Facebook’s
  satellite internet team joins Amazon}}, (Accessed: 2022-Jul-10) (2021).
\newline\urlprefix\url{https://www.theverge.com/2021/7/14/22576788/amazon-acquires-facebook-satellite-team-project-kuiper}

\bibitem{fang2021_iotj}
X.~Fang, W.~Feng, T.~Wei, Y.~Chen, N.~Ge, C.-X. Wang, {5G Embraces Satellites
  for 6G Ubiquitous IoT: Basic Models for Integrated Satellite Terrestrial
  Networks}, IEEE Internet of Things Journal 8~(18) (2021) 14399--14417.

\bibitem{Rappaport2019}
T.~S. {Rappaport}, Y.~{Xing}, O.~{Kanhere}, S.~{Ju}, A.~{Madanayake},
  S.~{Mandal}, A.~{Alkhateeb}, G.~C. {Trichopoulos}, {Wireless Communications
  and Applications Above 100 GHz: Opportunities and Challenges for 6G and
  Beyond}, IEEE Access 7 (2019) 78729--78757.

\bibitem{santamarta_2014_blackhat}
R.~Santamarta, {SATCOM terminals: Hacking by air, sea, and land}, Blackhat USA.

\bibitem{sat_under_attack}
J.~Trevithick,
  \href{https://www.thedrive.com/the-war-zone/43328/u-s-satellites-are-being-attacked-everyday-according-to-space-force-general}{{U.S.
  Satellites Are Being Attacked Every Day According To Space Force General}},
  (Accessed: 2022-Jul-10) (2021).
\newline\urlprefix\url{https://www.thedrive.com/the-war-zone/43328/u-s-satellites-are-being-attacked-everyday-according-to-space-force-general}

\bibitem{manulis2020cyber}
M.~Manulis, C.~Bridges, R.~Harrison, V.~Sekar, A.~Davis, {Cyber security in New
  Space: Analysis of threats, key enabling technologies and challenges},
  International Journal of Information Security (2020) 1--25.

\bibitem{Kodheli2021_comst}
O.~Kodheli, E.~Lagunas, N.~Maturo, S.~K. Sharma, B.~Shankar, J.~F.~M. Montoya,
  J.~C.~M. Duncan, D.~Spano, S.~Chatzinotas, S.~Kisseleff, J.~Querol, L.~Lei,
  T.~X. Vu, G.~Goussetis, {Satellite Communications in the New Space Era: A
  Survey and Future Challenges}, IEEE Communications Surveys \& Tutorials
  23~(1) (2021) 70--109.

\bibitem{Li2020_iotj}
B.~{Li}, Z.~{Fei}, C.~{Zhou}, Y.~{Zhang}, {Physical-Layer Security in Space
  Information Networks: A Survey}, IEEE Internet of Things Journal 7~(1) (2020)
  33--52.

\bibitem{zidan2020gnss}
J.~Zidan, E.~Adegoke, E.~Kampert, S.~A. Birrell, C.~R. Ford, M.~D. Higgins,
  Gnss vulnerabilities and existing solutions: A review of the literature, IEEE
  Access.

\bibitem{morales2019_comst}
R.~Morales-Ferre, P.~Richter, E.~Falletti, A.~de~la Fuente, E.~S. Lohan, {A
  Survey on Coping With Intentional Interference in Satellite Navigation for
  Manned and Unmanned Aircraft}, IEEE Communications Surveys \& Tutorials
  22~(1) (2019) 249--291.

\bibitem{rath2020security}
M.~Rath, S.~Mishra, {Security Approaches in Machine Learning for Satellite
  Communication}, in: Machine Learning and Data Mining in Aerospace Technology,
  Springer, 2020, pp. 189--204.

\bibitem{junzhi2019research}
L.~Junzhi, L.~Wanqing, F.~Qixiang, L.~Beidian, Research progress of gnss
  spoofing and spoofing detection technology, in: 2019 IEEE 19th International
  Conference on Communication Technology (ICCT), IEEE, 2019, pp. 1360--1369.

\bibitem{margaria_2017_ieeesignal}
D.~{Margaria}, B.~{Motella}, M.~{Anghileri}, J.~{Floch},
  I.~{Fernandez-Hernandez}, M.~{Paonni}, {Signal Structure-Based Authentication
  for Civil GNSSs: Recent Solutions and Perspectives}, IEEE Signal Processing
  Magazine 34~(5) (2017) 27--37.

\bibitem{saroj2016survey}
T.~Saroj, G.~S. Gaba, S.~K. Arora, {A survey on authentication schemes for
  satellite communications}, International Journal of Control Theory and
  Applications.

\bibitem{radhakrishnan2016survey}
R.~Radhakrishnan, W.~W. Edmonson, F.~Afghah, R.~M. Rodriguez-Osorio, F.~Pinto,
  S.~C. Burleigh, Survey of inter-satellite communication for small satellite
  systems: Physical layer to network layer view, IEEE Communications Surveys \&
  Tutorials 18~(4) (2016) 2442--2473.

\bibitem{schmidt2016survey}
D.~Schmidt, K.~Radke, S.~Camtepe, E.~Foo, M.~Ren, {A survey and analysis of the
  GNSS spoofing threat and countermeasures}, ACM Computing Surveys (CSUR)
  48~(4) (2016) 1--31.

\bibitem{hosseinidehaj2018_comst}
N.~Hosseinidehaj, Z.~Babar, R.~Malaney, S.~X. Ng, L.~Hanzo, {Satellite-Based
  Continuous-Variable Quantum Communications: State-of-the-Art and a Predictive
  Outlook}, IEEE Communications Surveys \& Tutorials 21~(1) (2018) 881--919.

\bibitem{guo2022_comst}
H.~Guo, J.~Li, J.~Liu, N.~Tian, N.~Kato, {A Survey on Space-Air-Ground-Sea
  Integrated Network Security in 6G}, IEEE Communications Surveys \& Tutorials
  24~(1) (2022) 53--87.

\bibitem{xu2021_twc}
Y.~Xu, J.~Liu, Y.~Shen, X.~Jiang, Y.~Ji, N.~Shiratori, {QoS-Aware Secure
  Routing Design for Wireless Networks With Selfish Jammers}, IEEE Transactions
  on Wireless Communications 20~(8) (2021) 4902--4916.

\bibitem{xu2021_iotj}
Y.~Xu, J.~Liu, Y.~Shen, J.~Liu, X.~Jiang, T.~Taleb, {Incentive Jamming-Based
  Secure Routing in Decentralized Internet of Things}, IEEE Internet of Things
  Journal 8~(4) (2021) 3000--3013.

\bibitem{maini2014}
A.~K. Maini, V.~Agrawal, {Satellite Orbits and Trajectories}, John Wiley \&
  Sons, 2014, pp. 37--78.

\bibitem{elbert2003_book}
B.~R. Elbert, {The Satellite Communication Applications Handbook (Artech House
  Space Applications Series)}, Artech House, Inc., USA, 2003.

\bibitem{cakaj2021_fcn}
S.~Cakaj, {The Parameters Comparison of the" Starlink" LEO Satellites
  Constellation for Different Orbital Shells}, Frontiers in Communications and
  Networks 2 (2021) 7.

\bibitem{peterson2003_satcom}
K.~M. Peterson, Satellite communications, in: R.~A. Meyers (Ed.), {Encyclopedia
  of Physical Science and Technology (Third Edition)}, third edition Edition,
  Academic Press, New York, 2003, pp. 413--438.

\bibitem{esa}
ESA,
  \href{https://www.esa.int/Applications/Telecommunications_Integrated_Applications/Satellite_frequency_bands}{{Satellite
  Frequency Bands}}, (Accessed: 2022-Jul-10) (2021).
\newline\urlprefix\url{https://www.esa.int/Applications/Telecommunications_Integrated_Applications/Satellite_frequency_bands}

\bibitem{Ilcev2018}
S.~D. Il{\v{c}}ev, Inmarsat GEO GMSC System, Springer International Publishing,
  Cham, 2018, pp. 1--100.

\bibitem{jeeff2019_spectrum}
J.~Foust, {SpaceX's space-Internet woes: Despite technical glitches, the
  company plans to launch the first of nearly 12,000 satellites in 2019}, IEEE
  Spectrum 56~(1) (2019) 50--51.

\bibitem{caprolu2020_commag}
M.~Caprolu, R.~D. Pietro, S.~Raponi, S.~Sciancalepore, P.~Tedeschi, {Vessels
  Cybersecurity: Issues, Challenges, and the Road Ahead}, IEEE Communications
  Magazine 58~(6) (2020) 90--96.

\bibitem{oligeri2020_wisec}
G.~Oligeri, S.~Sciancalepore, R.~Di~Pietro, {GNSS Spoofing Detection via
  Opportunistic IRIDIUM Signals}, in: Proceedings of the 13th ACM Conference on
  Security and Privacy in Wireless and Mobile Networks, WiSec '20, Association
  for Computing Machinery, New York, NY, USA, 2020, p. 42–52.

\bibitem{yuma2018_ieice}
Y.~Abe, H.~Tsuji, A.~Miura, S.~Adachi, {Frequency Resource Management Based on
  Model Predictive Control for Satellite Communications System}, {IEICE
  Transactions on Fundamentals of Electronics, Communications and Computer
  Sciences} E101.A~(12) (2018) 2434--2445.

\bibitem{lin2021_commag}
X.~Lin, S.~Rommer, S.~Euler, E.~A. Yavuz, R.~S. Karlsson, {5G from Space: An
  Overview of 3GPP Non-Terrestrial Networks}, IEEE Communications Standards
  Magazine 5~(4) (2021) 147--153.

\bibitem{aptsky}
S.~Tanase,
  \href{https://securelist.com/satellite-turla-apt-command-and-control-in-the-sky/72081/}{{Satellite
  Turla: APT Command and Control in the Sky}}, (Accessed: 2022-Jul-10) (2015).
\newline\urlprefix\url{https://securelist.com/satellite-turla-apt-command-and-control-in-the-sky/72081/}

\bibitem{mukherjee2014_comst}
A.~Mukherjee, S.~A.~A. Fakoorian, J.~Huang, A.~L. Swindlehurst, {Principles of
  physical layer security in multiuser wireless networks: A survey}, IEEE
  Communications Surveys \& Tutorials 16~(3) (2014) 1550--1573.

\bibitem{an2016_jsac}
K.~{An}, M.~{Lin}, J.~{Ouyang}, W.~{Zhu}, {Secure Transmission in Cognitive
  Satellite Terrestrial Networks}, IEEE Journal on Selected Areas in
  Communications 34~(11) (2016) 3025--3037.

\bibitem{lei_2011_tifs}
J.~{Lei}, Z.~{Han}, M.~A. {Vazquez-Castro}, A.~{Hjorungnes}, {Secure Satellite
  Communication Systems Design With Individual Secrecy Rate Constraints}, IEEE
  Transactions on Information Forensics and Security 6~(3) (2011) 661--671.

\bibitem{zheng_2012_twc}
G.~{Zheng}, P.~{Arapoglou}, B.~{Ottersten}, {Physical Layer Security in
  Multibeam Satellite Systems}, IEEE Transactions on Wireless Communications
  11~(2) (2012) 852--863.

\bibitem{xu_2019_ieeecomlet}
R.~{Xu}, X.~{Da}, H.~{Hu}, L.~{Ni}, Y.~{Pan}, {Self-Interference Cancellation
  Scheme for Secure AF Satellite Communication Based on FH-MWFRFT}, IEEE
  Communications Letters 23~(11) (2019) 2050--2053.

\bibitem{luo_2017_access}
Z.~{Luo}, H.~{Wang}, K.~{Zhou}, {Polarization Filtering Based Physical-Layer
  Secure Transmission Scheme for Dual-Polarized Satellite Communication}, IEEE
  Access 5 (2017) 24706--24715.

\bibitem{lu_2019_taes}
W.~{Lu}, K.~{An}, T.~{Liang}, {Robust Beamforming Design for Sum Secrecy Rate
  Maximization in Multibeam Satellite Systems}, IEEE Transactions on Aerospace
  and Electronic Systems 55~(3) (2019) 1568--1572.

\bibitem{xu_2019_ietcomm}
R.~Xu, X.~Da, Y.~Liang, L.~Ni, H.~Hu, {Secure transmission in AF satellite
  system based on FH-MWFRFT and null space beamforming}, IET Communications
  13~(10) (2019) 1506--1513.

\bibitem{lin_2018_jsac}
M.~{Lin}, Z.~{Lin}, W.~{Zhu}, J.~{Wang}, {Joint Beamforming for Secure
  Communication in Cognitive Satellite Terrestrial Networks}, IEEE Journal on
  Selected Areas in Communications 36~(5) (2018) 1017--1029.

\bibitem{zhang_2017_access}
X.~{Zhang}, B.~{Zhang}, D.~{Guo}, {Physical Layer Secure Transmission Based on
  Fast Dual Polarization Hopping in Fixed Satellite Communication}, IEEE Access
  5 (2017) 11782--11790.

\bibitem{lin_2018_jsac2}
Z.~{Lin}, M.~{Lin}, J.~{Wang}, Y.~{Huang}, W.~{Zhu}, {Robust Secure Beamforming
  for 5G Cellular Networks Coexisting With Satellite Networks}, IEEE Journal on
  Selected Areas in Communications 36~(4) (2018) 932--945.

\bibitem{kalantari_2015_tifs}
A.~{Kalantari}, G.~{Zheng}, Z.~{Gao}, Z.~{Han}, B.~{Ottersten}, {Secrecy
  Analysis on Network Coding in Bidirectional Multibeam Satellite
  Communications}, IEEE Transactions on Information Forensics and Security
  10~(9) (2015) 1862--1874.

\bibitem{huang_2017_ietcomm}
Q.~{Huang}, M.~{Lin}, K.~{An}, J.~{Ouyang}, W.~{Zhu}, {Secrecy performance of
  hybrid satellite-terrestrial relay networks in the presence of multiple
  eavesdroppers}, IET Communications 12~(1) (2017) 26--34.

\bibitem{yan_2016_wocc}
Y.~{Yan}, {Bangning Zhang}, {Daoxing Guo}, {Shengnan Li}, {Hehao Niu}, {Xi
  Wang}, {Joint beamforming and jamming design for secure cooperative hybrid
  satellite-terrestrial relay network}, in: 2016 25th Wireless and Optical
  Communication Conference (WOCC), 2016, pp. 1--5.

\bibitem{liu_2018_cse}
J.~{Liu}, J.~{Wang}, W.~{Liu}, Q.~{Wang}, M.~{Wang}, {A novel cooperative
  physical layer security scheme for satellite downlinks}, Chinese Journal of
  Electronics 27~(4) (2018) 860--865.

\bibitem{an_2016_wocc2}
K.~{An}, M.~{Lin}, {Tao Liang}, {Jian Ouyang}, {Can Yuan}, {Weixin Lu},
  {Secrecy performance analysis of land mobile satellite communication systems
  over Shadowed-Rician fading channels}, in: 2016 25th Wireless and Optical
  Communication Conference (WOCC), 2016, pp. 1--4.

\bibitem{li_2018_ieeewcl}
B.~{Li}, Z.~{Fei}, X.~{Xu}, Z.~{Chu}, {Resource Allocations for Secure
  Cognitive Satellite-Terrestrial Networks}, IEEE Wireless Communications
  Letters 7~(1) (2018) 78--81.

\bibitem{li_2018_tvt}
B.~{Li}, Z.~{Fei}, Z.~{Chu}, F.~{Zhou}, K.~{Wong}, P.~{Xiao}, {Robust
  Chance-Constrained Secure Transmission for Cognitive Satellite–Terrestrial
  Networks}, IEEE Transactions on Vehicular Technology 67~(5) (2018)
  4208--4219.

\bibitem{guo_2019_access}
K.~{Guo}, K.~{An}, Y.~{Huang}, B.~{Zhang}, {Physical Layer Security of
  Multiuser Satellite Communication Systems With Channel Estimation Error and
  Multiple Eavesdroppers}, IEEE Access 7 (2019) 96253--96262.

\bibitem{xu_2019_access}
R.~{Xu}, X.~{Da}, H.~{Hu}, L.~{Ni}, Y.~{Pan}, {A Secure Hybrid
  Satellite-Terrestrial Communication Network With AF/DF and Relay Selection},
  IEEE Access 7 (2019) 171980--171994.

\bibitem{knopp2021}
M.~G. Schraml, R.~T. Schwarz, A.~Knopp, {Multiuser MIMO Concept for Physical
  Layer Security in Multibeam Satellite Systems}, {IEEE Transactions on
  Information Forensics and Security} 16 (2021) 1670--1680.

\bibitem{Aliberti2016}
G.~Aliberti, R.~D. Pietro, S.~Guarino, {Reliable and perfectly secret
  communication over the generalized Ozarow-Wyner’s wire-tap channel},
  Computer Networks 109 (2016) 21 -- 30, special issue on Recent Advances in
  Physical-Layer Security.

\bibitem{barros2006secrecy}
J.~Barros, M.~R. Rodrigues, {Secrecy Capacity of Wireless Channels}, in: 2006
  IEEE international symposium on information theory, IEEE, 2006, pp. 356--360.

\bibitem{tedeschi2020_cst}
P.~Tedeschi, S.~Sciancalepore, R.~Di~Pietro, {Security in Energy Harvesting
  Networks: A Survey of Current Solutions and Research Challenges}, IEEE
  Communications Surveys \& Tutorials 22~(4) (2020) 2658--2693.

\bibitem{sciancalepore2018_cns}
S.~{Sciancalepore}, G.~{Oligeri}, R.~D. {Pietro}, {Shooting to the Stars:
  Secure Location Verification via Meteor Burst Communications}, in: 2018 IEEE
  Conference on Communications and Network Security (CNS), 2018, pp. 1--9.

\bibitem{wu2020_access}
Z.~Wu, Y.~Zhang, Y.~Yang, C.~Liang, R.~Liu, {Spoofing and Anti-Spoofing
  Technologies of Global Navigation Satellite System: A Survey}, IEEE Access 8
  (2020) 165444--165496.

\bibitem{gps_sdr_sim}
Osqzss, \href{https://github.com/osqzss/gps-sdr-sim}{{GPS-SDR-SIM}}, (Accessed:
  2022-Jul-10) (2021).
\newline\urlprefix\url{https://github.com/osqzss/gps-sdr-sim}

\bibitem{oligeri2019_wisec}
{G. Oligeri, S. Sciancalepore, O. Ibrahim, et al.}, {Drive Me Not: GPS Spoofing
  Detection via Cellular Network: (Architectures, Models, and Experiments)},
  in: Proceedings of the 12th Conference on Security and Privacy in Wireless
  and Mobile Networks, WiSec ’19, 2019, p. 12–22.

\bibitem{humphreys_2013_taes}
T.~E. {Humphreys}, {Detection Strategy for Cryptographic GNSS Anti-Spoofing},
  IEEE Transactions on Aerospace and Electronic Systems 49~(2) (2013)
  1073--1090.

\bibitem{psiaki_2013_iongnssm}
M.~{Psiaki}, S.~{Powell}, B.~{O'hanlon}, {GNSS spoofing detection using
  high-frequency antenna motion and carrier-phase data}, in: Proceedings of the
  ION GNSS+ Meeting, 2013, pp. 2949--2991.

\bibitem{ieee_2013_taes}
D.~{Borio}, {PANOVA Tests and their Application to GNSS Spoofing Detection},
  IEEE Transactions on Aerospace and Electronic Systems 49~(1) (2013) 381--394.

\bibitem{wesson_2013_gcsip}
K.~D. {Wesson}, B.~L. {Evans}, T.~E. {Humphreys}, {A combined symmetric
  difference and power monitoring GNSS anti-spoofing technique}, in: 2013 IEEE
  Global Conference on Signal and Information Processing, 2013, pp. 217--220.

\bibitem{heng2014gps}
L.~Heng, D.~B. Work, G.~X. Gao, Gps signal authentication from cooperative
  peers, IEEE Transactions on Intelligent Transportation Systems 16~(4) (2014)
  1794--1805.

\bibitem{yu2014short}
D.-Y. Yu, A.~Ranganathan, T.~Locher, S.~Capkun, D.~Basin, {Short paper:
  Detection of GPS spoofing attacks in power grids}, in: Proceedings of the
  2014 ACM conference on Security and privacy in wireless \& mobile networks,
  2014, pp. 99--104.

\bibitem{heng2013cooperative}
L.~Heng, D.~B. Work, G.~Gao, {Cooperative GNSS authentication}, Reliability
  from unreliable peers. Inside GNSS 8 (2013) 70--75.

\bibitem{psiaki2014gnss}
M.~L. Psiaki, B.~W. O'hanlon, S.~P. Powell, J.~A. Bhatti, K.~D. Wesson, T.~E.
  Humphreys, {GNSS spoofing detection using two-antenna differential carrier
  phase}, in: Radionavigation Laboratory Conference Proceedings, 2014.

\bibitem{stenberg2020gnss}
N.~Stenberg, E.~Axell, J.~Rantakokko, G.~Hendeby, {GNSS Spoofing Mitigation
  Using Multiple Receivers}, in: 2020 IEEE/ION Position, Location and
  Navigation Symposium (PLANS), IEEE, 2020, pp. 555--565.

\bibitem{o2013real}
B.~W. O'Hanlon, M.~L. Psiaki, J.~A. Bhatti, D.~P. Shepard, T.~E. Humphreys,
  {Real-time GPS spoofing detection via correlation of encrypted signals},
  Navigation 60~(4) (2013) 267--278.

\bibitem{hu2018gnss}
Y.~Hu, S.~Bian, K.~Cao, B.~Ji, {GNSS spoofing detection based on new signal
  quality assessment model}, GPS Solutions 22~(1) (2018) 1--13.

\bibitem{bhamidipati2018gps}
S.~Bhamidipati, T.~Y. Mina, G.~X. Gao, {GPS time authentication against
  spoofing via a network of receivers for power systems}, in: 2018 IEEE/ION
  Position, Location and Navigation Symposium (PLANS), IEEE, 2018, pp.
  1485--1491.

\bibitem{mina2018detecting}
T.~Y. Mina, S.~Bhamidipati, G.~X. Gao, Detecting gps spoofing via a
  multi-receiver hybrid communication network for power grid timing
  verification, in: Proceedings of the 31st International Technical Meeting of
  the Satellite Division of The Institute of Navigation (ION GNSS+ 2018), Hyatt
  Regency Miami, FL, USA, 2018, pp. 24--28.

\bibitem{hu2018gnss2}
Y.~Hu, S.~Bian, B.~Ji, J.~Li, {GNSS spoofing detection technique using fraction
  parts of double-difference carrier phases}, The Journal of Navigation 71~(5)
  (2018) 1111--1129.

\bibitem{formaggio2018authentication}
F.~Formaggio, S.~Tomasin, G.~Caparra, S.~Ceccato, N.~Laurenti, {Authentication
  of Galileo GNSS Signal by Superimposed Signature with Artificial Noise}, in:
  2018 26th European Signal Processing Conference (EUSIPCO), IEEE, 2018, pp.
  2573--2577.

\bibitem{formaggio2019authentication}
F.~Formaggio, S.~Tomasin, Authentication of satellite navigation signals by
  wiretap coding and artificial noise, EURASIP Journal on Wireless
  Communications and Networking 2019~(1) (2019) 1--17.

\bibitem{schmidt2020gps}
E.~Schmidt, N.~Gatsis, D.~Akopian, {A GPS spoofing detection and classification
  correlator-based technique using the LASSO}, IEEE Transactions on Aerospace
  and Electronic Systems.

\bibitem{anderson2017chips}
J.~M. Anderson, K.~L. Carroll, N.~P. DeVilbiss, J.~T. Gillis, J.~C. Hinks,
  B.~W. O’Hanlon, J.~J. Rushanan, L.~Scott, R.~A. Yazdi, {Chips-message
  robust authentication (Chimera) for GPS civilian signals}, in: Proceedings of
  the 30th International Technical Meeting of The Satellite Division of the
  Institute of Navigation (ION GNSS+ 2017), 2017, pp. 2388--2416.

\bibitem{tohidi2020effective}
S.~Tohidi, M.~R. Mosavi, {Effective Detection of GNSS Spoofing Attack Using A
  Multi-Layer Perceptron Neural Network Classifier Trained by PSO}, in: 2020
  25th International Computer Conference, Computer Society of Iran (CSICC),
  IEEE, 2020, pp. 1--5.

\bibitem{gross2018maximum}
J.~N. Gross, C.~Kilic, T.~E. Humphreys, {Maximum-likelihood power-distortion
  monitoring for GNSS-signal authentication}, IEEE Transactions on Aerospace
  and Electronic Systems 55~(1) (2018) 469--475.

\bibitem{wang2017gnss}
F.~Wang, H.~Li, M.~Lu, {GNSS spoofing detection and mitigation based on maximum
  likelihood estimation}, Sensors 17~(7) (2017) 1532.

\bibitem{wesson2017gnss}
K.~D. Wesson, J.~N. Gross, T.~E. Humphreys, B.~L. Evans, Gnss signal
  authentication via power and distortion monitoring, IEEE Transactions on
  Aerospace and Electronic Systems 54~(2) (2017) 739--754.

\bibitem{jiang2018satellite}
Y.~Jiang, Y.~Xing, {Satellite Spoofing Identification Method Based on Radio
  Frequency Feature Extraction}, JPhCS 1069~(1) (2018) 012079.

\bibitem{zou2016detection}
Q.~Zou, S.~Huang, F.~Lin, M.~Cong, {Detection of GPS spoofing based on UAV
  model estimation}, in: IECON 2016-42nd Annual Conference of the IEEE
  Industrial Electronics Society, IEEE, 2016, pp. 6097--6102.

\bibitem{axell2015gnss}
E.~Axell, E.~G. Larsson, D.~Persson, {GNSS spoofing detection using multiple
  mobile COTS receivers}, in: 2015 IEEE International Conference on Acoustics,
  Speech and Signal Processing (ICASSP), IEEE, 2015, pp. 3192--3196.

\bibitem{caparra2016autonomous}
G.~Caparra, C.~Wullems, R.~T. Ioannides, {An autonomous GNSS anti-spoofing
  technique}, in: 2016 8th ESA Workshop on Satellite Navigation Technologies
  and European Workshop on GNSS Signals and Signal Processing (NAVITEC), IEEE,
  2016, pp. 1--8.

\bibitem{singh2020mitigating}
S.~Singh, J.~Singh, S.~Singh, {Mitigating Spoofed GNSS Trajectories through
  Nature Inspired Algorithm}, GeoInformatica (2020) 1--20.

\bibitem{semanjski2020gnss}
S.~Semanjski, I.~Semanjski, W.~De~Wilde, S.~Gautama, {GNSS Spoofing Detection
  by Supervised Machine Learning with Validation on Real-World Meaconing and
  Spoofing Data—Part II}, Sensors 20~(7) (2020) 1806.

\bibitem{falco2019algorithm}
G.~Falco, M.~Nicola, E.~Falletti, M.~Pini, {An algorithm for finding the
  direction of arrival of counterfeit GNSS signals on a civil aircraft}, in:
  Proceedings of the 32nd International Technical Meeting of the Satellite
  Division of The Institute of Navigation (ION GNSS+ 2019), 2019, pp.
  3185--3196.

\bibitem{vadlamani2016}
S.~Vadlamani, B.~Eksioglu, H.~Medal, A.~Nandi, {Jamming attacks on wireless
  networks: A taxonomic survey}, International Journal of Production Economics
  172 (2016) 76--94.

\bibitem{zhang_tsp}
L.~{Zhang}, J.~{Ren}, T.~{Li}, {Time-Varying Jamming Modeling and
  Classification}, {IEEE Transactions on Signal Processing} 60~(7) (2012)
  3902--3907.

\bibitem{wang2018adaptive}
T.~Wang, X.~Wei, J.~Fan, T.~Liang, {Adaptive jammer localization in wireless
  networks}, Computer Networks 141 (2018) 17--30.

\bibitem{morehouse2021incremental}
T.~Morehouse, C.~Montes, M.~Bisbano, J.~F. Lin, M.~Shao, R.~Zhou, {Incremental
  learning-based jammer classification}, in: Artificial Intelligence and
  Machine Learning for Multi-Domain Operations Applications III, Vol. 11746,
  International Society for Optics and Photonics, 2021, p. 117462E.

\bibitem{borio2016}
D.~Borio, {Swept GNSS jamming mitigation through pulse blanking}, in: 2016
  European Navigation Conference (ENC), 2016, pp. 1--8.

\bibitem{topal2020}
O.~A. Topal, S.~Gecgel, E.~M. Eksioglu, G.~{Karabulut Kurt}, {Identification of
  smart jammers: Learning-based approaches using wavelet preprocessing},
  Physical Communication 39 (2020) 101029.

\bibitem{dempster_2016_ieee}
A.~G. {Dempster}, E.~{Cetin}, {Interference Localization for Satellite
  Navigation Systems}, Proceedings of the IEEE 104~(6) (2016) 1318--1326.

\bibitem{borio_2016_ieee}
D.~{Borio}, F.~{Dovis}, H.~{Kuusniemi}, L.~{Lo Presti}, {Impact and Detection
  of GNSS Jammers on Consumer Grade Satellite Navigation Receivers},
  Proceedings of the IEEE 104~(6) (2016) 1233--1245.

\bibitem{jung2018_taes}
H.~Jung, K.~Kim, J.~Kang, T.~S. Lee, S.~Kim, {An iALM-ICA-Based Antijamming
  DS-CDMA Receiver for LMS Systems}, IEEE Transactions on Aerospace and
  Electronic Systems 54~(5) (2018) 2318--2328.
\newblock \href {http://dx.doi.org/10.1109/TAES.2018.2814319}
  {\path{doi:10.1109/TAES.2018.2814319}}.

\bibitem{tamazin2020robust}
M.~Tamazin, A.~Noureldin, {Robust GPS Anti-jamming Technique Based on Fast
  Orthogonal Search}, in: Recent Advances in Engineering Mathematics and
  Physics, Springer, 2020, pp. 233--244.

\bibitem{bavzec2016gps}
M.~Ba{\v{z}}ec, B.~Luin, F.~Dimc, {GPS jamming detection with SDR}, in: Proc.
  of the 24th International Symposium on Electronics in Transport (ISEP 2016),
  ITS for efficient energy use, Electrotechnical Association of Slovenia,
  Ljubljana, Slovenia, Mar, 2016, pp. 1--4.

\bibitem{purwar2016gps}
A.~Purwar, D.~Joshi, V.~K. Chaubey, {GPS signal jamming and anti-jamming
  strategy—A theoretical analysis}, in: 2016 IEEE Annual India Conference
  (INDICON), IEEE, 2016, pp. 1--6.

\bibitem{gao2017gnss}
P.~Gao, S.~Sun, Z.~Zeng, C.~Wang, {GNSS spoofing jamming recognition based on
  machine learning}, in: International Conference On Signal And Information
  Processing, Networking And Computers, Springer, 2017, pp. 221--228.

\bibitem{glomsvoll2017gnss}
O.~Glomsvoll, L.~K. Bonenberg, {GNSS jamming resilience for close to shore
  navigation in the Northern Sea}, The Journal of Navigation 70~(1) (2017)
  33--48.

\bibitem{gao_2016_ieee}
G.~X. {Gao}, M.~{Sgammini}, M.~{Lu}, N.~{Kubo}, {Protecting GNSS Receivers From
  Jamming and Interference}, Proceedings of the IEEE 104~(6) (2016) 1327--1338.

\bibitem{lichtman_2016_ijscn}
M.~{Lichtman}, J.~{Reed}, {Analysis of reactive jamming against satellite
  communications}, International Journal of Satellite Communications and
  Networking 34~(2) (2016) 195--210.

\bibitem{borio_2015_iclgnss}
D.~{Borio}, C.~{Gioia}, {Real-time jamming detection using the sum-of-squares
  paradigm}, in: 2015 International Conference on Localization and GNSS
  (ICL-GNSS), 2015, pp. 1--6.

\bibitem{shi_2016_cns}
Y.~{Shi}, Y.~E. {Sagduyu}, {Spectrum learning and access for cognitive
  satellite communications under jamming}, in: 2016 IEEE Conference on
  Communications and Network Security (CNS), 2016, pp. 472--479.

\bibitem{borio_2016_enc}
D.~{Borio}, {Swept GNSS jamming mitigation through pulse blanking}, in: 2016
  European Navigation Conference (ENC), 2016, pp. 1--8.

\bibitem{wang_2017_milcom}
Q.~{Wang}, T.~{Nguyen}, K.~{Pham}, H.~{Kwon}, {Satellite jamming: A game
  theoretic analysis}, in: MILCOM 2017 - 2017 IEEE Military Communications
  Conference (MILCOM), 2017, pp. 141--146.

\bibitem{lang_2017_plos}
R.~{Lang}, H.~{Xiao}, Z.~{Li}, L.~{Yu}, {A anti-jamming method for satellite
  navigation system based on multi-objective optimization technique}, PloS one
  12~(7).

\bibitem{wu_2017_isspit}
Z.~{Wu}, Y.~{Zhao}, Z.~{Yin}, H.~{Luo}, {Jamming signals classification using
  convolutional neural network}, in: 2017 IEEE International Symposium on
  Signal Processing and Information Technology (ISSPIT), 2017, pp. 62--67.

\bibitem{sun_2015_cisp}
L.~{Sun}, B.~{Jing}, Y.~{Cheng}, L.~{Yuan}, {Jamming monitoring and
  anti-jamming by polarization diversity reception in the satellite navigation
  system}, in: 2015 8th International Congress on Image and Signal Processing
  (CISP), 2015, pp. 1303--1307.

\bibitem{wang_2017_iaces}
H.~{Wang}, L.~{Yang}, Y.~{Yang}, H.~{Zhang}, {Anti-jamming of Beidou navigation
  based on polarization sensitive array}, in: 2017 International Applied
  Computational Electromagnetics Society Symposium (ACES), 2017, pp. 1--2.

\bibitem{dong_2017_icwcsp}
K.~{Dong}, Z.~{Zhang}, X.~{Xu}, {A hybrid interference suppression scheme for
  global navigation satellite systems}, in: 2017 9th International Conference
  on Wireless Communications and Signal Processing (WCSP), 2017, pp. 1--7.

\bibitem{wang_2019_taes}
P.~{Wang}, Y.~{Wang}, E.~{Cetin}, A.~G. {Dempster}, S.~{Wu}, {GNSS Jamming
  Mitigation Using Adaptive-Partitioned Subspace Projection Technique}, IEEE
  Transactions on Aerospace and Electronic Systems 55~(1) (2019) 343--355.

\bibitem{hannon_2016_milcom}
M.~{Hannon}, {Shaung Feng}, {Hyuck Kwon}, {Khanh Pham}, {Jamming
  statistics-dependent frequency hopping}, in: MILCOM 2016 - 2016 IEEE Military
  Communications Conference, 2016, pp. 138--143.

\bibitem{winter_2016_milcom}
S.~P. {Winter}, C.~A. {Hofmann}, A.~{Knopp}, {Antenna diversity techniques for
  enhanced jamming resistance in multi-beam satellites}, in: MILCOM 2016 - 2016
  IEEE Military Communications Conference, 2016, pp. 618--623.

\bibitem{lubbers_2015_iainwc}
B.~{Lubbers}, S.~{Mildner}, P.~{Oonincx}, A.~{Scheele}, {A study on the
  accuracy of GPS positioning during jamming}, in: 2015 International
  Association of Institutes of Navigation World Congress (IAIN), 2015, pp.
  1--6.

\bibitem{chien_2017_tfeccs}
Y.~{Chien}, P.~{Chen}, S.~{Fang}, {Novel anti-jamming algorithm for GNSS
  receivers using wavelet-packet-transform-based adaptive predictors}, IEICE
  Transactions on fundamentals of electronics, communications and computer
  sciences 100~(2) (2017) 602--610.

\bibitem{hurley2007}
C.~Hurley, R.~Rogers, F.~Thornton, D.~Connelly, B.~Baker, {Chapter 2 -
  Understanding Antennas and Antenna Theory}, in: C.~Hurley, R.~Rogers,
  F.~Thornton, D.~Connelly, B.~Baker (Eds.), {WarDriving and Wireless
  Penetration Testing}, Syngress, Rockland, 2007, pp. 31--61.

\bibitem{viasat}
ESA,
  \href{https://www.viasat.com/space-innovation/space-systems/intersatellite-communications/}{{ViaSat
  - Intersatellite communications}}, (Accessed: 2022-Jul-10) (2021).
\newline\urlprefix\url{https://www.viasat.com/space-innovation/space-systems/intersatellite-communications/}

\bibitem{gong2020_comst}
S.~Gong, X.~Lu, D.~T. Hoang, D.~Niyato, L.~Shu, D.~I. Kim, Y.-C. Liang, {Toward
  smart wireless communications via intelligent reflecting surfaces: A
  contemporary survey}, IEEE Communications Surveys \& Tutorials 22~(4) (2020)
  2283--2314.

\bibitem{almohamad2020_ojcs}
A.~Almohamad, A.~M. Tahir, A.~Al-Kababji, H.~M. Furqan, T.~Khattab, M.~O.
  Hasna, H.~Arslan, {Smart and secure wireless communications via reflecting
  intelligent surfaces: A short survey}, IEEE Open Journal of the
  Communications Society 1 (2020) 1442--1456.

\bibitem{dong2021_icc}
H.~Dong, C.~Hua, L.~Liu, W.~Xu, {Towards Integrated Terrestrial-Satellite
  Network via Intelligent Reflecting Surface}, in: ICC 2021 - IEEE
  International Conference on Communications, 2021, pp. 1--6.
\newblock \href {http://dx.doi.org/10.1109/ICC42927.2021.9500640}
  {\path{doi:10.1109/ICC42927.2021.9500640}}.

\bibitem{xu2021_tvt}
S.~Xu, J.~Liu, Y.~Cao, J.~Li, Y.~Zhang, {Intelligent reflecting surface enabled
  secure cooperative transmission for satellite-terrestrial integrated
  networks}, IEEE Transactions on Vehicular Technology 70~(2) (2021)
  2007--2011.

\bibitem{tekbiyik2020_arxiv}
K.~Tekb{\i}y{\i}k, G.~K. Kurt, A.~R. Ekti, A.~G{\"o}r{\c{c}}in,
  H.~Yanikomeroglu, {Reconfigurable intelligent surfaces empowered THz
  communication in LEO satellite networks}, arXiv preprint arXiv:2007.04281.

\bibitem{simeuri2021_icl}
A.~Siemuri, H.~Kuusniemi, M.~S. Elmusrati, P.~Välisuo, A.~Shamsuzzoha,
  {Machine Learning Utilization in GNSS—Use Cases, Challenges and Future
  Applications}, in: 2021 International Conference on Localization and GNSS
  (ICL-GNSS), 2021, pp. 1--6.
\newblock \href {http://dx.doi.org/10.1109/ICL-GNSS51451.2021.9452295}
  {\path{doi:10.1109/ICL-GNSS51451.2021.9452295}}.

\bibitem{calvo2020_wowmom}
R.~Calvo-Palomino, A.~Bhattacharya, G.~Bovet, D.~Giustiniano, {Short:
  LSTM-based GNSS Spoofing Detection Using Low-cost Spectrum Sensors}, in: 2020
  IEEE 21st International Symposium on "A World of Wireless, Mobile and
  Multimedia Networks" (WoWMoM), 2020, pp. 273--276.
\newblock \href {http://dx.doi.org/10.1109/WoWMoM49955.2020.00055}
  {\path{doi:10.1109/WoWMoM49955.2020.00055}}.

\bibitem{semanjski2019_icl}
S.~Semanjski, A.~Muls, I.~Semanjski, W.~De~Wilde, {Use and Validation of
  Supervised Machine Learning Approach for Detection of GNSS Signal Spoofing},
  in: 2019 International Conference on Localization and GNSS (ICL-GNSS), 2019,
  pp. 1--6.
\newblock \href {http://dx.doi.org/10.1109/ICL-GNSS.2019.8752775}
  {\path{doi:10.1109/ICL-GNSS.2019.8752775}}.

\bibitem{semanjski2020_sensors}
S.~Semanjski, I.~Semanjski, W.~De~Wilde, A.~Muls, {Use of supervised machine
  learning for GNSS signal spoofing detection with validation on real-world
  meaconing and spoofing data—Part I}, Sensors 20~(4) (2020) 1171.

\bibitem{ibrahim_2016_scn}
M.~H. {Ibrahim}, S.~{Kumari}, A.~K. {Das}, V.~{Odelu}, {Jamming resistant
  non-interactive anonymous and unlinkable authentication scheme for mobile
  satellite networks}, Security and Communication Networks 9~(18) (2016)
  5563--5580.

\bibitem{chen_2014_amis}
C.~L. Chen, K.~W. Cheng, Y.~L. Chen, C.~Chang, C.~C. Lee, {An improvement on
  the self-verification authentication mechanism for a mobile satellite
  communication system}, Applied Mathematics \& Information Sciences 8~(1L)
  (2014) 97--106.

\bibitem{xinghua_2017_isc}
W.~{Xinghua}, Z.~{Aixin}, L.~{Jianhua}, Z.~{Weiwei}, L.~{Yuchen}, {A
  Lightweight Authentication and Key Agreement Scheme for Mobile Satellite
  Communication Systems}, in: Information Security and Cryptology, Springer
  International Publishing, Cham, 2017, pp. 187--204.

\bibitem{xu_2020_ijscn}
S.~{Xu}, X.~{Liu}, M.~{Ma}, J.~{Chen}, {An improved mutual authentication
  protocol based on perfect forward secrecy for satellite communications},
  International Journal of Satellite Communications and Networking 38~(1)
  (2020) 62--73.

\bibitem{zhang_2015_ijscn}
Y.~{Zhang}, J.~{Chen}, B.~{Huang}, {An improved authentication scheme for
  mobile satellite communication systems}, International Journal of Satellite
  Communications and Networking 33~(2) (2015) 135--146.

\bibitem{lin_2016_ijscn}
H.~{Lin}, {Efficient dynamic authentication for mobile satellite communication
  systems without verification table}, International Journal of Satellite
  Communications and Networking 34~(1) (2016) 3--10.

\bibitem{zhao_2016_milcom}
W.~{Zhao}, A.~{Zhang}, J.~{Li}, X.~{Wu}, Y.~{Liu}, {Analysis and design of an
  authentication protocol for space information network}, in: MILCOM 2016 -
  2016 IEEE Military Communications Conference, 2016, pp. 43--48.

\bibitem{liu_2017_ijscn}
Y.~{Liu}, A.~{Zhang}, S.~{Li}, J.~{Tang}, J.~{Li}, {A lightweight
  authentication scheme based on self-updating strategy for space information
  network}, International Journal of Satellite Communications and Networking
  35~(3) (2017) 231--248.

\bibitem{caparra_2016_iclgnss}
G.~{Caparra}, S.~{Sturaro}, N.~{Laurenti}, C.~{Wullems}, {Evaluating the
  security of one-way key chains in TESLA-based GNSS Navigation Message
  Authentication schemes}, in: 2016 International Conference on Localization
  and GNSS (ICL-GNSS), 2016, pp. 1--6.

\bibitem{caparra2016novel}
G.~Caparra, S.~Sturaro, N.~Laurenti, C.~Wullems, R.~T. Ioannides, {A novel
  navigation message authentication scheme for GNSS open service}, in: ION
  GNSS, Vol. 2016, 2016.

\bibitem{hernandez_2014_ion-gnss}
I.~F. Hern{\'a}ndez, V.~Rijmen, G.~S. Granados, J.~Sim{\'o}n,
  I.~Rodr{\'\i}guez, J.~D. Calle, {Design drivers, solutions and robustness
  assessment of navigation message authentication for the galileo open
  service}, in: Proceedings of the 27th international technical meeting of the
  satellite division of the institute of navigation (ION GNSS 2014), 2014, pp.
  2810--2827.

\bibitem{kerns2014blueprint}
A.~J. Kerns, K.~D. Wesson, T.~E. Humphreys, A blueprint for civil gps
  navigation message authentication, in: 2014 IEEE/ION Position, Location and
  Navigation Symposium-PLANS 2014, IEEE, 2014, pp. 262--269.

\bibitem{huang2020mutual}
C.~Huang, Z.~Zhang, M.~Li, L.~Zhu, Z.~Zhu, X.~Yang, A mutual authentication and
  key update protocol in satellite communication network, Automatika 61~(3)
  (2020) 334--344.

\bibitem{jurcut2019novel}
A.~D. Jurcut, J.~Chen, A.~Kalla, M.~Liyanage, J.~Murphy, {A Novel
  Authentication Mechanism for Mobile Satellite Communication Systems}, in:
  2019 IEEE Wireless Communications and Networking Conference Workshop (WCNCW),
  IEEE, 2019, pp. 1--7.

\bibitem{ghorbani2020navigation}
K.~Ghorbani, N.~Orouji, M.~Mosavi, {Navigation Message Authentication Based on
  One-Way Hash Chain to Mitigate Spoofing Attacks for GPS L1}, Wireless
  Personal Communications 113~(4) (2020) 1743--1754.

\bibitem{curran_2014_encgnss}
J.~T. Curran, M.~Paonni, J.~Bishop, {Securing the open-service: A candidate
  navigation message authentication scheme for galileo E1 OS}, in: European
  Navigation Conference,(ENC-GNSS), 2014.

\bibitem{meng2018low}
W.~Meng, K.~Xue, J.~Xu, J.~Hong, N.~Yu, Low-latency authentication against
  satellite compromising for space information network, in: 2018 IEEE 15th
  International Conference on Mobile Ad Hoc and Sensor Systems (MASS), IEEE,
  2018, pp. 237--244.

\bibitem{perrig2002tesla}
A.~Perrig, R.~Canetti, J.~D. Tygar, D.~Song, {The TESLA broadcast
  authentication protocol}, {Rsa Cryptobytes} 5~(2) (2002) 2--13.

\bibitem{proverif}
B.~Blanchet, {Automatic Verification of Correspondences for Security
  Protocols}, Journal of Computer Security 17~(4) (2009) 363--434.

\bibitem{blanchet2017cryptoverif}
B.~Blanchet, {CryptoVerif: A computationally-sound security protocol verifier},
  Tech. Rep.

\bibitem{avispa}
A.~Armando, D.~Basin, Y.~Boichut, Y.~Chevalier, L.~Compagna, J.~Cuellar, P.~H.
  Drielsma, P.~C. He{\'a}m, O.~Kouchnarenko, J.~Mantovani, S.~M{\"o}dersheim,
  D.~von Oheimb, M.~Rusinowitch, J.~Santiago, M.~Turuani, L.~Vigan{\`o},
  L.~Vigneron, The avispa tool for the automated validation of internet
  security protocols and applications, in: K.~Etessami, S.~K. Rajamani (Eds.),
  Computer Aided Verification, Springer Berlin Heidelberg, Berlin, Heidelberg,
  2005, pp. 281--285.

\bibitem{tamarin}
S.~Meier, B.~Schmidt, C.~Cremers, D.~Basin, The tamarin prover for the symbolic
  analysis of security protocols, in: N.~Sharygina, H.~Veith (Eds.), Computer
  Aided Verification, Springer Berlin Heidelberg, Berlin, Heidelberg, 2013, pp.
  696--701.

\bibitem{stallings2012_book}
W.~Stallings, L.~Brown, M.~D. Bauer, A.~K. Bhattacharjee, {Computer Security:
  Principles and Practice}, Pearson Education Upper Saddle River, NJ, USA,
  2012.

\bibitem{ostadsharif_2019_comcom}
A.~{Ostad-Sharif}, D.~{Abbasinezhad-Mood}, M.~{Nikooghadam}, {Efficient
  utilization of elliptic curve cryptography in design of a three-factor
  authentication protocol for satellite communications}, Computer
  Communications 147 (2019) 85 -- 97.

\bibitem{murtaza2019lightweight}
A.~Murtaza, T.~Xu, S.~Jahanzeb, H.~Pirzada, L.~Jianwei, {A lightweight
  authentication and key sharing protocol for satellite communication}, Int. J.
  Comput. Commun. Control (2019, in press).

\bibitem{caparra2017key}
G.~Caparra, S.~Ceccato, S.~Sturaro, N.~Laurenti, {A key management architecture
  for GNSS open service Navigation Message Authentication}, in: 2017 European
  navigation conference (ENC), IEEE, 2017, pp. 287--297.

\bibitem{deng_2018_iaeac}
L.~{Deng}, S.~{Ye}, H.~{Qiu}, {Transmission Security Platform for
  Transportation Information based on BeiDou Navigation Satellite System}, in:
  2018 IEEE 3rd Advanced Information Technology, Electronic and Automation
  Control Conference (IAEAC), 2018, pp. 2110--2113.

\bibitem{altaf2020lightweight}
I.~Altaf, M.~A. Saleem, K.~Mahmood, S.~Kumari, P.~Chaudhary, C.-M. Chen, {A
  Lightweight Key Agreement and Authentication Scheme for
  Satellite-Communication Systems}, IEEE Access 8 (2020) 46278--46287.

\bibitem{yantao_2010_jcn}
Z.~{Yantao}, M.~{Jianfeng}, {A highly secure identity-based authenticated
  key-exchange protocol for satellite communication}, Journal of Communications
  and Networks 12~(6) (2010) 592--599.

\bibitem{lee_2013_ijscn}
C.-C. Lee, {A simple key agreement scheme based on chaotic maps for VSAT
  satellite communications}, International journal of satellite communications
  and networking 31~(4) (2013) 177--186.

\bibitem{qi_2019_ijscn}
M.~{Qi}, J.~{Chen}, Y.~{Chen}, {A secure authentication with key agreement
  scheme using ECC for satellite communication systems}, International Journal
  of Satellite Communications and Networking 37~(3) (2019) 234--244.

\bibitem{joye2009_ide}
M.~Joye, G.~Neven, {Identity-based cryptography}, Vol.~2, IOS press, 2009.

\bibitem{kocarev2001_csm}
L.~Kocarev, {Chaos-based cryptography: a brief overview}, {IEEE Circuits and
  Systems Magazine} 1~(3) (2001) 6--21.

\bibitem{sarr2010_springer}
A.~P. Sarr, P.~Elbaz-Vincent, J.-C. Bajard, A new security model for
  authenticated key agreement, in: J.~A. Garay, R.~De~Prisco (Eds.), Security
  and Cryptography for Networks, Springer Berlin Heidelberg, Berlin,
  Heidelberg, 2010, pp. 219--234.

\bibitem{bellare1998modular}
M.~Bellare, R.~Canetti, H.~Krawczyk, {A modular approach to the design and
  analysis of authentication and key exchange protocols}, in: Proceedings of
  the thirtieth annual ACM symposium on Theory of computing, 1998, pp.
  419--428.

\bibitem{diamanti2016_qi}
E.~Diamanti, H.-K. Lo, B.~Qi, Z.~Yuan, {Practical challenges in quantum key
  distribution}, npj Quantum Information 2~(1) (2016) 1--12.

\bibitem{gyongyosi2019_csr}
L.~Gyongyosi, S.~Imre, {A Survey on quantum computing technology}, {Computer
  Science Review} 31 (2019) 51--71.

\bibitem{khan_2018_osa}
I.~{Khan}, B.~{Heim}, A.~{Neuzner}, C.~{Marquardt}, {Satellite-based QKD},
  Optics and Photonics News 29~(2) (2018) 26--33.

\bibitem{bedington_2017_nqi}
R.~{Bedington}, J.~M. {Arrazola}, A.~{Ling}, {Progress in satellite quantum key
  distribution}, npj Quantum Information 3~(1) (2017) 1--13.

\bibitem{benton_2010_oc}
D.~M. Benton, P.~M. Gorman, P.~R. Tapster, D.~M. Taylor, {A compact free space
  quantum key distribution system capable of daylight operation}, Optics
  communications 283~(11) (2010) 2465--2471.

\bibitem{tomaello_2011_asr}
A.~Tomaello, C.~Bonato, V.~Da~Deppo, G.~Naletto, P.~Villoresi, {Link budget and
  background noise for satellite quantum key distribution}, Advances in Space
  Research 47~(5) (2011) 802--810.

\bibitem{mafu_2013_pra}
M.~Mafu, A.~Dudley, S.~Goyal, D.~Giovannini, M.~McLaren, M.~J. Padgett,
  T.~Konrad, F.~Petruccione, N.~L{\"u}tkenhaus, A.~Forbes, Higher-dimensional
  orbital-angular-momentum-based quantum key distribution with mutually
  unbiased bases, Physical Review A 88~(3) (2013) 032305.

\bibitem{vallone2015experimental}
G.~Vallone, D.~Bacco, D.~Dequal, S.~Gaiarin, V.~Luceri, G.~Bianco,
  P.~Villoresi, {Experimental satellite quantum communications}, Physical
  Review Letters 115~(4) (2015) 040502.

\bibitem{tan_2015_jmo}
Y.~C. Tan, R.~Chandrasekara, C.~Cheng, A.~Ling, {Radiation tolerance of
  opto-electronic components proposed for space-based quantum key
  distribution}, Journal of Modern Optics 62~(20) (2015) 1709--1712.

\bibitem{bourgoin_2015_pra}
J.-P. Bourgoin, N.~Gigov, B.~L. Higgins, Z.~Yan, E.~Meyer-Scott, A.~K.
  Khandani, N.~L{\"u}tkenhaus, T.~Jennewein, {Experimental quantum key
  distribution with simulated ground-to-satellite photon losses and processing
  limitations}, Physical Review A 92~(5) (2015) 052339.

\bibitem{liao_2017_nature}
{S. {Liao} and W. {Cai} and W. {Liu} and L. {Zhang} and Y. {Li} and J. {Ren}
  and J. {Yin} and Q. {Shen} and Y. {Cao} and Z. {Li} et al},
  {Satellite-to-ground quantum key distribution}, Nature 549~(7670) (2017)
  43--47.

\bibitem{takenaka_2017_nature}
H.~Takenaka, A.~Carrasco-Casado, M.~Fujiwara, M.~Kitamura, M.~Sasaki,
  M.~Toyoshima, {Satellite-to-ground quantum-limited communication using a
  50-kg-class microsatellite}, Nature photonics 11~(8) (2017) 502--508.

\bibitem{toyoshima_2011_ijo}
M.~Toyoshima, H.~Takenaka, Y.~Shoji, Y.~Takayama, M.~Takeoka, M.~Fujiwara,
  M.~Sasaki, {Polarization-basis tracking scheme in satellite quantum key
  distribution}, International Journal of Optics 2011.

\bibitem{jennewein_2014_etsd}
T.~Jennewein, C.~Grant, E.~Choi, C.~Pugh, C.~Holloway, J.~Bourgoin, H.~Hakima,
  B.~Higgins, R.~Zee, {The NanoQEY mission: ground to space quantum key and
  entanglement distribution using a nanosatellite}, in: Emerging technologies
  in security and defence II; and quantum-physics-based information security
  III, Vol. 9254, International Society for Optics and Photonics, 2014, p.
  925402.

\bibitem{sharma_2018_icccnt}
V.~{Sharma}, S.~{Banerjee}, {Analysis of quantum key distribution based
  satellite communication}, in: 2018 9th International Conference on Computing,
  Communication and Networking Technologies (ICCCNT), IEEE, 2018, pp. 1--5.

\bibitem{bonato_2009_njp}
C.~Bonato, A.~Tomaello, V.~Da~Deppo, G.~Naletto, P.~Villoresi, {Feasibility of
  satellite quantum key distribution}, New Journal of Physics 11~(4) (2009)
  045017.

\bibitem{bedington_2015_small}
R.~Bedington, T.~Zhongkan, R.~Chandrasekara, C.~Cheng, T.~Y. Chuan, K.~Durak,
  A.~V. Zafra, E.~Truong-cao, A.~Ling, D.~Oi, {Small Photon Entangling Quantum
  System (SPEQS) Enabling Space Based Quantum Key Distribution (QKD)},
  International Astronautical Congress, Jerusalem, Israel.

\bibitem{wang_2013_nature}
J.-Y. Wang, B.~Yang, S.-K. Liao, L.~Zhang, Q.~Shen, X.-F. Hu, J.-C. Wu, S.-J.
  Yang, H.~Jiang, Y.-L. Tang, et~al., {Direct and full-scale experimental
  verifications towards ground--satellite quantum key distribution}, Nature
  Photonics 7~(5) (2013) 387--393.

\bibitem{liao_2017_nature2}
S.-K. Liao, H.-L. Yong, C.~Liu, G.-L. Shentu, D.-D. Li, J.~Lin, H.~Dai, S.-Q.
  Zhao, B.~Li, J.-Y. Guan, et~al., {Long-distance free-space quantum key
  distribution in daylight towards inter-satellite communication}, Nature
  Photonics 11~(8) (2017) 509--513.

\bibitem{rarity_2002_njp}
J.~G. Rarity, P.~Tapster, P.~Gorman, P.~Knight, {Ground to satellite secure key
  exchange using quantum cryptography}, New Journal of Physics 4~(1) (2002) 82.

\bibitem{chen2021nature}
Y.-A. Chen, Q.~Zhang, T.-Y. Chen, W.-Q. Cai, S.-K. Liao, J.~Zhang, K.~Chen,
  J.~Yin, J.-G. Ren, Z.~Chen, et~al., {An integrated space-to-ground quantum
  communication network over 4,600 kilometres}, {Nature} 589~(7841) (2021)
  214--219.

\bibitem{quantropi}
I.~Quantropi, \href{https://lp.quantropi.com/qispace-trial}{{Quantropi
  QiSpace}}, (Accessed: 2022-Jul-10) (2015).
\newline\urlprefix\url{https://lp.quantropi.com/qispace-trial}

\bibitem{takeda2021quantum}
K.~Takeda, A.~Noiri, T.~Nakajima, J.~Yoneda, T.~Kobayashi, S.~Tarucha, {Quantum
  tomography of an entangled three-qubit state in silicon}, {Nature
  Nanotechnology} 16~(9) (2021) 965--969.

\bibitem{miralem2021_csur}
M.~Mehic, M.~Niemiec, S.~Rass, J.~Ma, M.~Peev, A.~Aguado, V.~Martin,
  S.~Schauer, A.~Poppe, C.~Pacher, M.~Voznak, {Quantum Key Distribution: A
  Networking Perspective}, ACM Computing Surveys 53~(5).

\bibitem{sharma2021_ojcs}
P.~Sharma, A.~Agrawal, V.~Bhatia, S.~Prakash, A.~K. Mishra, {Quantum Key
  Distribution Secured Optical Networks: A Survey}, IEEE Open Journal of the
  Communications Society 2 (2021) 2049--2083.

\bibitem{xu2020_rmp}
F.~Xu, X.~Ma, Q.~Zhang, H.-K. Lo, J.-W. Pan, {Secure quantum key distribution
  with realistic devices}, Rev. Mod. Phys. 92 (2020) 025002.

\bibitem{li2021physrev}
W.~Li, V.~Zapatero, H.~Tan, K.~Wei, H.~Min, W.-Y. Liu, X.~Jiang, S.-K. Liao,
  C.-Z. Peng, M.~Curty, F.~Xu, J.-W. Pan, {Experimental Quantum Key
  Distribution Secure Against Malicious Devices}, Phys. Rev. Applied.

\bibitem{salahdine2020security}
F.~Salahdine, N.~Kaabouch, {Security threats, detection, and countermeasures
  for physical layer in cognitive radio networks: A survey}, Physical
  Communication 39 (2020) 101001.

\bibitem{mozaffari2019_comst}
{M. {Mozaffari}, et al.}, {A Tutorial on UAVs for Wireless Networks:
  Applications, Challenges, and Open Problems}, IEEE Communications Surveys \&
  Tutorials 21~(3).

\bibitem{yin2022_tits}
Z.~Yin, M.~Jia, N.~Cheng, W.~Wang, F.~Lyu, Q.~Guo, X.~Shen, {UAV-Assisted
  Physical Layer Security in Multi-Beam Satellite-Enabled Vehicle
  Communications}, {IEEE Transactions on Intelligent Transportation Systems}
  23~(3) (2022) 2739--2751.

\bibitem{cheng2019_jsac}
N.~Cheng, F.~Lyu, W.~Quan, C.~Zhou, H.~He, W.~Shi, X.~Shen,
  {Space/Aerial-Assisted Computing Offloading for IoT Applications: A
  Learning-Based Approach}, {IEEE Journal on Selected Areas in Communications}
  37~(5) (2019) 1117--1129.

\bibitem{zhang2017_commag}
N.~Zhang, S.~Zhang, P.~Yang, O.~Alhussein, W.~Zhuang, X.~S. Shen, {Software
  Defined Space-Air-Ground Integrated Vehicular Networks: Challenges and
  Solutions}, {IEEE Communications Magazine} 55~(7) (2017) 101--109.

\bibitem{emilien2021}
A.-V. Emilien, C.~Thomas, H.~Thomas, {UAV \& satellite synergies for optical
  remote sensing applications: A literature review}, Science of Remote Sensing
  3 (2021) 100019.

\bibitem{ray2021_ksu}
P.~P. Ray, {A review on 6G for space-air-ground integrated network: Key
  enablers, open challenges, and future direction}, {Journal of King Saud
  University - Computer and Information Sciences}.

\bibitem{oligeri2020past}
G.~Oligeri, S.~Raponi, S.~Sciancalepore, R.~Di~Pietro, {PAST-AI: Physical-layer
  Authentication of Satellite Transmitters via Deep Learning}, arXiv preprint
  arXiv:2010.05470.

\bibitem{rechenberg2021_icmcis}
M.~von Rechenberg, P.~H.~L. Rettore, R.~R.~F. Lopes, P.~Sevenich,
  {Software-Defined Networking Applied in Tactical Networks: Problems,
  Solutions and Open Issues}, in: 2021 International Conference on Military
  Communication and Information Systems (ICMCIS), 2021, pp. 1--8.

\bibitem{papa2020_tnsm}
A.~Papa, T.~de~Cola, P.~Vizarreta, M.~He, C.~Mas-Machuca, W.~Kellerer, {Design
  and Evaluation of Reconfigurable SDN LEO Constellations}, IEEE Transactions
  on Network and Service Management 17~(3) (2020) 1432--1445.

\bibitem{bertaux2015_commag}
L.~Bertaux, S.~Medjiah, P.~Berthou, S.~Abdellatif, A.~Hakiri, P.~Gelard,
  F.~Planchou, M.~Bruyere, {Software defined networking and virtualization for
  broadband satellite networks}, IEEE Communications Magazine 53~(3) (2015)
  54--60.

\bibitem{eliyan2021_fgcs}
L.~F. Eliyan, R.~{Di Pietro}, {DoS and DDoS attacks in Software Defined
  Networks: A survey of existing solutions and research challenges}, Future
  Generation Computer Systems 122 (2021) 149--171.

\bibitem{ahan2021_tnsm}
A.~Kak, I.~F. Akyildiz, {Towards Automatic Network Slicing for the Internet of
  Space Things}, IEEE Transactions on Network and Service Management (2021)
  1--1.

\bibitem{akyildiz2019_ieeenet}
I.~F. Akyildiz, A.~Kak, {The Internet of Space Things/CubeSats}, IEEE Network
  33~(5) (2019) 212--218.

\bibitem{akyildiz2019_comnet}
I.~F. Akyildiz, A.~Kak, {The Internet of Space Things/CubeSats: A ubiquitous
  cyber-physical system for the connected world}, Computer Networks 150 (2019)
  134--149.

\bibitem{greensat}
ESA,
  \href{https://blogs.esa.int/cleanspace/2016/10/24/how-do-you-build-a-green-satellite/}{{How
  do you build a Green Satellite?}}, (Accessed: 2022-Jul-10) (2021).
\newline\urlprefix\url{https://blogs.esa.int/cleanspace/2016/10/24/how-do-you-build-a-green-satellite/}

\bibitem{kassas2021_taes}
Z.~Z. Kassas, J.~Khalife, M.~Neinavaie, {The First Carrier Phase Tracking and
  Positioning Results with Starlink LEO Satellite Signals}, IEEE Transactions
  on Aerospace and Electronic Systems (2021) 1--1.

\bibitem{Scholl2021}
M.~Scholl, (draft) introduction to cybersecurity for commercial satellite
  operations, National Institute of Standards and Technology ({NIST}).

\bibitem{rinaldi2020_access}
F.~Rinaldi, H.-L. Maattanen, J.~Torsner, S.~Pizzi, S.~Andreev, A.~Iera,
  Y.~Koucheryavy, G.~Araniti, {Non-Terrestrial Networks in 5G \& Beyond: A
  Survey}, IEEE Access 8 (2020) 165178--165200.
\newblock \href {http://dx.doi.org/10.1109/ACCESS.2020.3022981}
  {\path{doi:10.1109/ACCESS.2020.3022981}}.

\bibitem{3gpp_rel16}
3GPP, \href{https://www.3gpp.org/ftp/Specs/archive/38_series/38.821/}{{TR
  38.821: Solutions for NR to support non-terrestrial networks (NTN).}},
  (Accessed: 2022-Jul-10) (2021).
\newline\urlprefix\url{https://www.3gpp.org/ftp/Specs/archive/38_series/38.821/}

\bibitem{3gpp_rel17}
3GPP, \href{https://www.3gpp.org/release-17}{{3GPP Release 17}}, (Accessed:
  2022-Jul-10) (2022).
\newline\urlprefix\url{https://www.3gpp.org/release-17}

\bibitem{3gpp_rel18}
3GPP, \href{https://www.3gpp.org/release18}{{3GPP Release 18}}, (Accessed:
  2022-Jul-10) (2022).
\newline\urlprefix\url{https://www.3gpp.org/release18}

\bibitem{kota2021_icc}
S.~Kota, G.~Giambene, {6G integrated non-terrestrial networks: Emerging
  technologies and challenges}, in: IEEE International Conference on
  Communications Workshops (ICC Workshops), IEEE, 2021, pp. 1--6.

\bibitem{deliv_sat5g}
SAT5G,
  \href{https://www.sat5g-project.eu/wp-content/uploads/2019/04/761413_Deliverable_25_Roadmap-for-Satellite-into-5G.pdf}{{Satellite
  and Terrestrial Network for 5G - D6.1: Roadmap to Satellite into 5G}},
  (Accessed: 2022-Jul-10) (2021).
\newline\urlprefix\url{https://www.sat5g-project.eu/wp-content/uploads/2019/04/761413_Deliverable_25_Roadmap-for-Satellite-into-5G.pdf}

\bibitem{wang2020_dcn}
M.~Wang, T.~Zhu, T.~Zhang, J.~Zhang, S.~Yu, W.~Zhou, {Security and privacy in
  6G networks: New areas and new challenges}, Digital Communications and
  Networks 6~(3) (2020) 281--291.

\bibitem{nguyen2021_comst}
V.-L. Nguyen, P.-C. Lin, B.-C. Cheng, R.-H. Hwang, Y.-D. Lin, {Security and
  Privacy for 6G: A Survey on Prospective Technologies and Challenges}, IEEE
  Communications Surveys \& Tutorials 23~(4) (2021) 2384--2428.

\bibitem{rose2020zero}
S.~Rose, O.~Borchert, S.~Mitchell, S.~Connelly, {Zero trust architecture},
  Tech. rep., {National Institute of Standards and Technology} (2020).

\end{thebibliography}

\end{document}